\documentclass[11pt,a4paper]{article}
\pdfoutput=1

\RequirePackage{ifpdf} 
\usepackage{amsmath} 
\usepackage{mathtools}

\usepackage{jheppub}
\usepackage{pstricks}
\usepackage[final]{pdfpages} 
\usepackage{ifpdf} 
\usepackage{slashed}
\usepackage{bbm}
\usepackage{soul}
\usepackage{hyperref}
\usepackage{cleveref}
\usepackage{comment}

\usepackage{color} 
\usepackage{graphics}

\usepackage{etoolbox} 
\usepackage{fixmath}

\usepackage{notoccite} 

\usepackage{caption} 
\usepackage{subcaption} 
\usepackage{amsfonts}

\usepackage{multirow}
\usepackage{epstopdf}

\usepackage{relsize}
\usepackage{cancel}

\usepackage[normalem]{ulem}

\usepackage{tikz}
\usetikzlibrary{positioning,arrows}
\usetikzlibrary{decorations.pathmorphing}
\usetikzlibrary{decorations.markings}
\usetikzlibrary{shapes.geometric}
\tikzset{
    vector/.style={decorate, decoration={snake}, draw},
    provector/.style={decorate, decoration={snake,amplitude=2.5pt}, draw},
    antivector/.style={decorate, decoration={snake,amplitude=-2.5pt}, draw},
    fermion/.style={draw=black,
      postaction={decorate},decoration={markings,mark=at position .55
        with {\arrow[draw=black]{>}}}}, 
    fermionbar/.style={draw=black, postaction={decorate},
                       decoration={markings,mark=at position .55 with {\arrow[draw=black]{<}}}},
    fermionnoarrow/.style={draw=black},
    gluon/.style={decorate, draw=black,decoration={coil,amplitude=4pt, segment length=4pt}},
    scalar/.style={dashed,draw=black,
      postaction={decorate},decoration={markings,mark=at position .55
        with {\arrow[draw=black]{>}}}}, 
    scalarbar/.style={dashed,draw=black,
      postaction={decorate},decoration={markings,mark=at position .55
        with {\arrow[draw=black]{<}}}}, 
    scalarnoarrow/.style={dashed,draw=black},
    electron/.style={draw=black,
      postaction={decorate},decoration={markings,mark=at position .55
        with {\arrow[draw=black]{>}}}}, 
    bigvector/.style={decorate, decoration={snake,amplitude=4pt}, draw},
}

\def\be{\begin{equation}}
\def\ee{\end{equation}}
\def\nn{\nonumber\\}
\def\ep{\epsilon}
\def\T{\mathcal{T}}
\def\V{\mathcal{V}}
\excludecomment{strikeout}

\preprint{IMSc/2018/10/07, IPPP/18/96}

\title{Second order splitting functions and infrared safe cross sections in $\mathcal{N}=4$ SYM theory}

\author{Pulak Banerjee$^{a}$, Amlan Chakraborty$^{a}$, Prasanna K.\
  Dhani$^{a,b}$,  V. Ravindran$^{a}$ and Satyajit
  Seth$^{c}$} 

\affiliation{$^a$ The Institute of Mathematical Sciences, HBNI, Taramani,
  Chennai-600113, India \\ 
 $^b$ INFN, Sezione di Firenze, I-50019 Sesto Fiorentino, Florence, Italy\\
$^c$ 
Institute for Particle Physics Phenomenology, Department of Physics,
University of Durham, Durham, DH1 3LE, UK} 

\emailAdd{bpulak@imsc.res.in}
\emailAdd{amlanchak@imsc.res.in}
\emailAdd{prasannakumar.dhani@fi.infn.it}
\emailAdd{ravindra@imsc.res.in}
\emailAdd{satyajit.seth@durham.ac.uk}
\abstract{
We report our findings on  the perturbative structure of ${\cal N}=4$ supersymmetric Yang-Mills (SYM) theory
in the infrared sector by computing inclusive scattering cross sections of on-shell particles.
We use half-BPS, energy-momentum tensor and Konishi operators to produce singlet states
in the scattering processes to probe the soft and the collinear properties of the cross sections.  
By appropriately defining the infrared safe observables, we obtain collinear splitting functions
up to second order in the perturbation theory.  The splitting functions and the infrared finite cross sections  
demonstrate several interesting connections with those in the perturbative QCD.
We also determine the process independent soft distribution function up to third order in the perturbation theory and
show that it is universal {\it i.e.} independent of the operators as well as the external states. Interestingly, the soft
distribution function in ${\cal N}=4$ SYM theory matches exactly with the leading transcendental part of the
corresponding one in the QCD. This enables us to predict the third order soft plus virtual cross section for the production of the on-shell singlet states.
}
\keywords{$\mathcal{N}=4$ SYM theory, Infrared, Factorization, Soft plus virtual cross sections}
\begin{document}
\allowdisplaybreaks[4]
\unitlength1cm

\maketitle
\flushbottom



\newcommand{\dis}{}
\newcommand{\overbar}[1]{mkern-1.5mu\overline{\mkern-1.5mu#1\mkern-1.5mu}\mkern
1.5mu}

\newcommand{\B}{\color{blue}}
\newcommand{\R}{\color{red}}


\section{Introduction}
\label{sec:intro}
Perturbative results from Quantum Chromodynamics (QCD) play an important role in understanding
the physics of strong interactions.  Inclusive and differential cross sections 
computed using perturbative QCD not only helped to discover several of elementary particles of the Standard Model (SM) but also provided a laboratory to understand the field theoretical structure of 
non-abelian gauge theories. 
\begin{strikeout}
 \st{For example, both theoretical and experimental results from 
high energetic collision processes, such as the  deep-inelastic scattering and the Drell-Yan production 
provides the complete knowledge of the internal structure as well as the dynamics of hadrons
in terms of their constituents such as quarks and gluons.}  
\end{strikeout}
{ Scattering cross sections computed in high energetic collision processes such as the Drell-Yan \cite{Drell:1970wh} and the  deep-inelastic scattering processes can be expressed in terms of perturbatively computed partonic cross section, convoluted with the parton distribution functions (PDFs).  The partons refer to quarks and gluons and the PDFs describe the probabilities
of finding the partons in a bound state.}
\begin{strikeout}
\st{These scattering cross sections at high energies can be expressed in terms of the perturbatively 
calculable scatterings involving constituents of hadrons properly convoluted with parton distribution functions.}
\st{These constituents at high energies are light quarks and gluons often called partons 
and the corresponding PDFs describe their
probabilities to exist in the hadron.  Such a description of hadronic cross section goes by
the name parton model.}
\end{strikeout} 
 While the scattering of partons 
are calculable order by order in perturbative QCD (pQCD), the non-perturbative PDFs 
\begin{strikeout}
\st{depend on the long distance
part of the hadronic cross section and hence} 
\end{strikeout}
{ are process independent and can be
computed only by non-perturbative techniques.}
\begin{strikeout}
\st{The PDFs can be in principle computable using non-perturbative techniques.  However, due to
the complexity involved in the computations, they are fitted from the data from various high
energy scattering experiments.}
\end{strikeout}
However, the evolution of PDFs as functions of energy scale is controlled by pQCD
through Altarelli-Parisi (AP) \cite{Altarelli:1977zs} splitting functions.

The study of the perturbative series at different orders give a wealth of informations about 
the structure of various divergences such as ultraviolet (UV) and infrared (IR) divergences.
\begin{strikeout}
\st{These divergences appear in the Feynman diagrams through loop and phase space integrals 
at the intermediate stages of computations of hadronic observables  such as cross sections and
decay rates involving hadrons.  In particular, they show up when the parton level cross sections
are computed beyond the leading order in perturbation theory.}
\end{strikeout}
Computation of partonic cross sections beyond the leading order (LO) in pQCD
introduces these divergences and the origin of these singularities is due to loop and phase space integrations. The UV divergences arise due to the 
high energy modes of virtual particles in the loop while the IR divergences 
such as soft and collinear ones, come from gluons and light quarks respectively.
Only certain quantities like inclusive and differential cross sections, decay rates 
computed in pQCD can be measured in the scattering experiments.
They go by the name infrared safe observables.
In these observables, the soft divergences cancel among themselves between real emission and virtual 
diagrams at every order in perturbation theory, the collinear divergences 
from degenerate final states again cancel among themselves when they are appropriately summed.
Hence, for scatterings or decays  { where quarks and/or gluons are absent in the initial state}
the resultant observables are infrared safe.
If the incoming states 
{ contain quarks and/or gluons, there will be
initial state collinear singularities.} 
\begin{strikeout}
\st{scatter, which will bring 
initial state collinear
singularities making the latter sensitive to initial state collinear singularities.}
\end{strikeout}
Thanks to the existence of bound states 
and the factorisation properties of the initial state collinear singularities,
one can remove these singularities by appropriately redefining the PDFs. 
In other words, collinear unsafe parton level cross sections resulting from scatterings of 
initial light partonic states can be factorised 
into process independent kernels and collinear finite coefficient
functions order by order in pQCD.   
The kernels satisfy renormalisation group equations controlled by 
AP splitting functions \cite{Altarelli:1977zs},
which are known exactly up to third order in perturbation series 
\cite{Gross:1973ju,Georgi:1951sr,Altarelli:1977zs, Floratos:1977au, Floratos:1978ny, GonzalezArroyo:1979df,GonzalezArroyo:1979he,Curci:1980uw,Furmanski:1980cm,Floratos:1981hs,Hamberg:1991qt,Vogt:2004mw,Moch:2004pa}; the four loop counterparts in planar and large $n_f$ (number of flavours) limit were calculated 
in  \cite{Moch:2017uml, Davies:2016jie}.
Thus in QCD, the nature of UV and IR divergences and their cancellation
at cross section level have been studied in details and is quite well understood.
\begin{strikeout}
\st{This knowledge of QCD can guide us to investigate how the UV and IR singularities
appear in different types of quantum field theories. }
\end{strikeout}
{This knowledge of UV and IR singularities in QCD can guide us to investigate the 
divergence structure arising in different quantum field theoretic context.}
One of the interesting candidate to study is the  $\mathcal{N}=4$ supersymmetric Yang-Mills (SYM) theory.
Like QCD, it is a renormalizable gauge theory
in four dimensional Minkowski space. In addition to having all the symmetries of QCD ,
$\mathcal{N}=4$ SYM theory possesses 
 supersymmetry and conformal symmetry that make it
interesting to study. Although the study of cross sections in such a theory has no 
phenomenological implications, yet they can help us to understand 
the factorization properties of the IR singularities, 
the latter being useful to extract the AP kernels at each order 
in the perturbation theory. {\it One of the goals in this article is to compute 
the AP splitting functions up to two-loop order in the perturbation series 
from explicit calculation of certain inclusive cross sections in $\mathcal{N}=$ 4 SYM theory.}

The most widely studied quantities in $\mathcal{N}=$ 4 SYM theory are the on-shell amplitudes. 
Owing to the supersymmetric Ward identities \cite{Grisaru:1976vm}, the tree level 
on-shell identical-helicity amplitudes vanish~\cite{Elvang:2013cua}. In addition these on-shell amplitudes 
satisfy the Anti-de-Sitter/conformal field theory (AdS/CFT) conjecture \cite{Maldacena:1997re}
which relates the maximally SYM theory in four dimensions and gravity in five-dimensional anti-de Sitter space. Such a duality proposes that
quantities computed in a perturbative expansion in $\mathcal{N}=$ 4 SYM theory should add up to a simple expression, so that they can be related to the weakly coupled 
gravity. In other words, the perturbatively computed quantities should be related to one another 
in order to reduce to such simple expressions. This property of supersymmetric amplitudes has been extensively studied
in the works ~\cite{Anastasiou:2003kj,Eden:1999kh,Eden:2000mv, Eden:2000vb}.
The factorization property of the finite terms for $n$-point $m$-loop  
amplitudes in terms of one-loop counterparts was shown in the article ~\cite{Bern:2005iz}.
However this factorization property fails beyond two-loop five-point maximally helicity violating (MHV) amplitudes 
\cite{Bartels:2008ce,Bern:2008ap}.

Like on-shell amplitudes, form factors (FFs) of composite operators also contribute
to the scattering cross sections and provide important information about the 
IR structure of the gauge theories. The FFs are defined as the matrix elements of the 
composite operator between an off-shell initial state and on-shell final states.
The most widely studied composite operator in $\mathcal{N}=$ 4 SYM  theory is the half-BPS
operator, whose UV anomalous dimensions vanish to all orders in perturbation theory
~\cite{Dobrev:1985vh,Ferrara:1999zg,Ferrara:1999cw, Minwalla:1997ka,Rasmussen:2000ii}.
As a result the FFs of this composite operator look relatively simple. The first computation 
of a two-point FF up to two-loop order for the half-BPS operator was done by van Neervan 
\cite{vanNeerven:1985ja}. The three loop computation was done in ~\cite{Gehrmann:2011xn}
where the authors have shown an interesting connection between their results and the corresponding 
ones in non-supersymmetric SU(N) gauge theory containing $n_f$ fermions, with the following replacement 
of the color factors: $C_A = C_F = n_f = N$, where $C_A$, $C_F$ are the Casimir for the adjoint, fundamental representations respectively. Study of FFs of composite operators
also shed light on the ADS/CFT correspondence, see~\cite{Brandhuber:2010ad,Gehrmann:2011xn,Boels:2015yna,Brandhuber:2012vm,Brandhuber:2011tv,Bork:2011cj} for details.
Over the past few years calculation of FFs for non-BPS type composite operators, such as the Konishi 
~\cite{Konishi:1983hf} have also gained interest. However this operator is non-protected and hence
develop UV anomalous dimensions at each order in perturbation theory. In this regard, study of the FFs of
the Konishi operator in $\mathcal{N}=$ 4 SYM  theory helps to understand the IR structure in a more general way.
For computation of one-loop two-point, two-loop two-point and one-loop three-point FFs 
see~\cite{Nandan:2014oga}. In~\cite{Ahmed:2016vgl}, some of the authors of the present paper have presented the
three-loop two-point FF for the Konishi operator and also predicted up to $1/\epsilon$
pole  at four loops in $d$ ($=4+\epsilon$) dimensions. The two-loop three-point FF and their
finite remainders for the half-BPS~\cite{Brandhuber:2012vm} and the Konishi operator were recently calculated 
in \cite{Banerjee:2016kri}.
Several other results on $n$-point FFs of the Konishi operator are now available, see
~\cite{Wilhelm:2014qua,Loebbert:2015ova,Brandhuber:2016fni,Loebbert:2016xkw} for details.

The FFs of composite operators as well as the on-shell amplitudes offer a wide scope to
investigate the IR structure of quantum field theory. In QCD, the two-point FFs satisfy the
K+G equation ~\cite{Sudakov:1954sw,Mueller:1979ih, Collins:1980ih,Sen:1981sd}
and the IR structure of these quantities are already well understood.  The universal 
nature of IR singularities for a $n$-point QCD amplitude up to two-loop order
was predicted by Catani in ~\cite{Catani:1998bh}. It was then realized in
~\cite{Sterman:2002qn} that the above predictions are a consequence of the underlying factorization 
and resummation properties of the QCD amplitudes. Later on 
the generalisation of the results in ~\cite{Catani:1998bh} and ~\cite{Sterman:2002qn} 
in SU(N) gauge theory, at any loop order, having $n_f$ light flavours in terms of cusp, collinear and soft anomalous dimensions
was formulated by Becher and Neubert ~\cite{Becher:2009cu} and independently by Gardi and Magnea 
~\cite{Gardi:2009qi}.  All these studies have helped to understand the iterative structure of IR divergences
which subsequently lead to the program of resummation of observables, the latter being an important 
area of study at the energies of the hadron colliders.

Undoubtedly, higher order computation of the FFs and the amplitudes unravel the IR structure of the $\mathcal{N}=4$ SYM
theory in an elegant way.
However purely real emission processes, which appear in cross sections, can also 
give important informations about the nature of soft and collinear emissions.  In QCD, 
the gluons in a virtual loop can become
soft and contribute to poles in $\epsilon$ in a dimensionally regulated theory, similar 
situation also happens when gluons in a real emission process carry a small fraction of the  momentum
of the incoming particles. More precisely, when we perform the phase space integrations for such real
emission processes, we encounter poles in $\epsilon$, at every order in perturbation series. 
These soft contributions from real and virtual diagrams
cancel order  by order when they are added together, thanks to the Kinoshita-Lee-Nauenberg (KLN) theorem \cite{Kinoshita:1962ur, Lee:1964is}.  In addition, the real emissions of gluons and quarks are sensitive to collinear singularities; 
while the final state divergences are
taken care by the KLN theorem, the initial state counterparts are removed by mass factorization.
Similar scattering of  massless gluons, quarks, scalars and pseudo-scalars in ${\cal N}=4$ SYM theory can be studied within
a supersymmetric preserving regularised scheme. 
The cancellation of soft singularities and factorisation of collinear singularities in the scattering
cross sections will also provide wealth of information on the IR structure of $\cal N$ = 4 gauge theory.
One can investigate the soft plus virtual part of these finite cross sections after mass factorisation
in terms of universal cusp and collinear anomalous dimensions. Also, the factorisation of 
initial state collinear
singularities provides valuable information about the AP splitting functions in $\cal N$=4 SYM theory.
Understanding such cross sections in the light of well known results in QCD will help us to
investigate the  resummation of soft gluon contributions to all orders in perturbation theory in a process
independent manner.  In other words, $\cal N$=4 SYM theory offers an easier framework to appreciate IR structure of not only on-shell amplitudes but also scattering cross sections. 
Such an exercise helps us to appreciate better the underlying principles of quantum field theory. 
In this article, we make such an attempt to
compute inclusive cross sections for the production of various singlet states through
effective interactions of    
certain composite operators, namely  the half-BPS, the Konishi and energy momentum (EM) tensor
with fields of ${\cal N}$=4 SYM theory.  In contrast to the
half-BPS and the Konishi operator, the EM tensor couples universally to all the fields; 
thus the number 
of processes contributing becomes overwhelmingly large. We compute all the subprocesses contributing up to 
two-loop order in the perturbation theory and use them to extract the AP kernels up to
the same order in perturbative expansion. We notice interesting aspects of the 
splitting functions, namely, presence of transcendental 
terms ranging $2 l$ ($l$ = loop order) to 0. We also compare the cross sections calculated in $\mathcal{N}=4$ SYM theory to the 
standard model counterparts, namely Drell-Yan and Higgs boson productions and find 
interesting similarities and differences, which we shall elucidate in the later part of the paper 
in detail.

The paper is organised as follows.  In Sec.~\ref{theory}, we start with the Lagrangian in ${\cal N}=4$ SYM theory, the interaction
of its fields with different external currents through composite operators and describe the general framework to compute
the collinear splitting kernels and infrared safe cross sections.  Sec.~\ref{method} contains the methodology to compute
scattering cross sections using the regularised version of the Lagrangian. In Sec.~\ref{results}, we present the results 
for the splitting functions and the coefficient functions up to two-loop level and discuss our findings 
in detail.  Finally Sec.~\ref{conclusion} is devoted to conclusions.   Appendix contains the Mellin-$j$ space results
of AP splitting functions in a compact form.

\section{Theoretical framework}
\label{theory}
\subsection{Lagrangian}
In this section we present the theoretical framework necessary for our computation. 
Our main interest is to understand the infrared structure of ${\cal N}=4$ SYM theory which will
eventually lead us to compute the AP splitting functions and find out many interesting aspects of the partonic cross sections.
We achieve this by computing inclusive cross sections for the production of various 
singlet states from the scattering of pair of on-shell particles belonging to ${\cal N}=4$ SYM theory.  
The effective Lagrangian responsible for the production of such states can come from
singlet composite operators such as the half-BPS, the Konishi and the EM tensor of ${\cal N}=4$ SYM theory.
These are denoted by ${\cal O}^I$ 
with $I =$ half-BPS,~${\cal K} \,\rm{and}\,T$ {being the three singlet operators} respectively. 
The Lagrangian density including the effective interactions reads as follows
\begin{align}
\mathcal{L} = \mathcal{L}^{{\cal N} = 4}_{\text{SYM}} + \mathcal{L}_{\text{int}},
\end{align}
where  $ \mathcal{L}^{{\cal N} = 4}_{\text{SYM}}$  ~\cite{Brink:1976bc,Gliozzi:1976qd, Jones:1977zr,Poggio:1977ma} in four space-time dimensions is given by 
\begin{align}
\label{Lagdens}
\mathcal{L}^{{\cal N} = 4}_{\text{SYM}} = &-\frac{1}{4}G_{\mu\nu}^a G^{\mu\nu a} - \frac{1}{2\xi}(\partial_{\mu}A^{a\mu})^2 + \partial_{\mu}\bar{\eta}^a 
D^{\mu}\eta_a + \frac{i}{2}\bar{\lambda}^a_m\gamma^{\mu}D_{\mu}\lambda^a_m + \frac{1}{2}(D_{\mu}\phi^a_i)^2 
\nonumber\\
&+ \frac{1}{2}(D_{\mu}\chi^a_i)^2 - \frac{g}{2}f^{abc}\bar{\lambda}^a_m[\alpha^i_{m,n}\phi^b_i +
 \gamma_5\beta^i_{m,n}\chi^b_i]\lambda^c_n - \frac{g^2}{4}\Big[(f^{abc}\phi^b_i\phi^c_j)^2 
\nonumber\\
&+ (f^{abc}\chi^b_i\chi^c_j)^2 + 2 (f^{abc}\phi^b_i\chi^c_j)^2\Big].
\end{align}
The fields $A^{a\mu}$ and $\eta^a$  represent the gauge and ghost fields respectively.
The Majorana fields are denoted by $\lambda^a_m$, with $m=1,...,4$ denoting their generation type.
The scalar and pseudoscalar fields are $ \phi^a_i$ and $\chi^a_j$ where indices  $i,\,j =1,2,3$ represent 
different types of scalars and  pseudoscalars in the theory.
All the fields transform in adjoint representation and hence carry  SU(N) color indices $a$.
$g$ is the coupling constant and $\xi$ is the gauge fixing parameter.
The gluonic field strength tensor is given by
$G_{\mu\nu}^a = \partial_{\mu}A_{\nu}^a - \partial_{\nu}A_{\mu}^a + g f^{abc}A_{\mu}^b A_{\nu}^c$,
while  covariant derivative  is  $D_{\mu} = \partial_{\mu} - i g T^a A_{\mu}^a$. 
The matrices $T^a$ satisfy
$[T^a,T^b]_{-} = i f^{abc}T^c$,
where $f^{abc}$ is the totally antisymmetric structure constant of the group algebra. The generators are normalized as
$\text{Tr}(T^a\,T^b) = \frac{1}{2}\,\delta^{a\,b}$. The six antisymmetric matrices $\alpha$ and $\beta$
satisfy the relations
\begin{equation}
\label{theory4}
[\alpha^i, \alpha^j]_{+} = [\beta^i, \beta^j]_{+} = -2 \delta^{ij} \mathbb{I},\quad  [\alpha^i, \beta^j]_{-} = 0,
\end{equation}
and in addition, 
\begin{equation}
\label{theory5}
\text{tr}(\alpha^i) = \text{tr}(\beta^i) = \text{tr}(\alpha^i\beta^j)= 0,\quad \text{tr}(\alpha^i\alpha^j) = \text{tr}(\beta^i\beta^j) = -4 \delta^{ij}.
\end{equation}
Since we work with the dimensionally regulated version of the Lagrangian density in the 
$d=4+\epsilon$ space-time dimensions, and use supersymmetry preserving
dimensional reduction scheme~\cite{Siegel:1979wq,Capper:1979ns},
the number of $\alpha$ and $\beta$ matrices is  dependent on  $d$.  Hence care is needed when
we perform the contraction of indices, for example      
\begin{equation}
\label{theory6}
\alpha^i\alpha^i = \beta^i\beta^i = \Big(-3 +  \frac{\epsilon}{2}\Big) \mathbb{I},
\quad\alpha^i\alpha^j\alpha^i  = \alpha^j \Big(1- \frac{\epsilon}{2}\Big) \mathbb{I},
\quad \beta^i\beta^j\beta^i  = \beta^j \Big(1- \frac{\epsilon}{2} \Big)  \mathbb{I}.
\end{equation}
The interaction part of the Lagrangian density in Eq.~(\ref{Lagdens}) is given by
\begin{equation}
 \mathcal{L}_{\text{int}} = \mathcal{L}^{\text{BPS}} + \mathcal{L}^{{\cal K}} + \mathcal{L}^{{\text{T}}},
\end{equation}
where 
\begin{equation}
\label{intlag}
\mathcal{L}^{\text{BPS}} = J^{\text{BPS}}_{rt}{\mathcal O}^{\text{BPS}}_{rt}, \quad  \mathcal{L}^{{\cal K}}
= J^{{\cal K}}{ \mathcal O}^{{\cal K}}, \quad \mathcal{L}^{{\text{T}}} = J^{\text{T}\mu \nu } \mathcal{O}^{\rm T}_{\mu \nu}.
\end{equation}
{ In the above, the different singlet states are denoted by external currents $J$s 
(namely $J^{\text{BPS}}_{rt}$,  $J^{{\cal K}}$ and $J_{\mu \nu}^{\rm{T}}$)
which couple to a half-BPS (${\mathcal O}^{\text{BPS}}_{rt}$), a Konishi (${ \mathcal O}^{{\cal K}}$) and 
a tensorial operator ($\mathcal{O}^{\rm T}_{\mu \nu}$).}
\begin{strikeout}
\st{(namely $J^{\text{BPS}}_{rt}$,  $J^{{\cal K}}$ and $J_{\mu \nu}^{T}$) which couple 
to a half-BPS, a Konishi and a tensorial operators respectively. }
\end{strikeout}
The half-BPS operator that we use is given by~\cite{Bergshoeff:1980is,vanNeerven:1985ja}
\begin{equation}
\label{Op1}
{\cal O}^{\rm{BPS}}_{rt} = \phi^a_r\phi^a_t - \frac{1}{3}\delta_{rt}\phi^a_s\phi^a_s.
\end{equation}
The factor 1/3 has been used to ensure the tracelessness property in four dimensions.    
The primary operator of the Konishi supermultiplet, the Konishi, has the following form
 \begin{equation}
 \label{Op2}
{\cal O}^{\cal K} = \phi^a_r\phi^a_r + \chi^a_r\chi^a_r .
 \end{equation} 
In terms of the Majorana, gauge, scalar and pseudoscalar fields, we find the EM tensor as  
\begin{align}
\label{eq:Tmunu}
\mathcal{O}^{\rm T}_{\mu \nu} = \,& G^{a}_{\mu \lambda} G^{\lambda}_{a \nu} + \frac{1}{4} 
\eta_{\mu\nu} G^{a}_{\rho \lambda} G^{\rho \lambda}_{a} - \frac{1}{\xi}\, \partial_{\lambda} A^{\lambda}
\left[ \partial_\mu A_\nu + \partial_\nu A_\mu  \right] + \frac{1}{2\xi}\, \eta_{\mu\nu} {(\partial_\rho {A^\rho}_a)}^2
\nn &
+ (\partial_\mu \bar{\eta}^{a}) (D_{\nu} \eta_{a}) + (\partial_\nu \bar{\eta}^{a}) (D_{\mu} \eta_{a})
 - \eta_{\mu \nu} (\partial_\rho \bar{\eta}^{a}) (D^{\rho} \eta_{a})
 + \frac{i}{4} \Big[\bar{\lambda}^{a}_{m} \gamma_{\mu} D_{\nu} \lambda^{a}_{m} 
 \nn & 
 + \bar{\lambda}^{a}_{m} \gamma_{\nu} 
 D_{\mu} \lambda^{a}_{m} - \frac{1}{2} \partial_{\mu} (
 \bar{\lambda}^{a}_{m} \gamma_{\nu} \lambda^{a}_{m})   
  -\frac{1}{2}  \partial_{\nu}   (\bar{\lambda}^{a}_{m} \gamma_{\mu} \lambda^{a}_{m} )\Big] 
  - \frac{i}{2} \Big[ \eta_{\mu \nu} \bar{\lambda}^{a}_{m} \gamma^{\rho} D_{\rho} \lambda^{a}_{m} 
  \nn &
  - \frac{1}{2} \eta_{\mu \nu} \partial_{\rho} \left( \bar{\lambda}^{a}_{m} \gamma^{\rho} \lambda^{a}_{m} \right) \Big]
  + (D_\mu \phi_{i}^{a}) (D_\nu \phi_{i}^{a}) - \frac{1}{2} \eta_{\mu \nu} (D_\rho \phi_{i}^{a})^{2}
   + (D_\mu \chi_{i}^{a}) (D_\nu \chi_{i}^{a})
 \nn &
 - \frac{1}{2} \eta_{\mu \nu} (D_\rho \chi_{i}^{a})^{2}  
+ \frac{g}{2} \eta_{\mu \nu} f^{abc} \,
 \bar{\lambda}_{m}^{a} \left[ \alpha^{i}_{m,n} \phi_{i}^{b} + \gamma_5 \beta^{i}_{m,n} \chi_{i}^{b}\right] \lambda_{n}^{c}
+\frac{g^2}{4} \eta_{\mu \nu} \Big[ (f^{abc} \phi^{b}_{i} \phi^{c}_{j})^2 
\nn &
+  (f^{abc} \chi^{b}_{i} \chi^{c}_{j})^2  
+ 2 (f^{abc} \phi^{b}_{i} \chi^{c}_{j} )^2\Big].
 \end{align}
In the next section, we will  evaluate the inclusive cross sections for the production of various singlet
states, $I$, { due to the interaction of the fields of $\mathcal{N}=4$ SYM theory. }
\begin{strikeout}
 \st{ which we hereafter call the half-BPS, Konishi and $T$ states.}
\end{strikeout}
\subsection{Computation of splitting functions and finite cross sections}
In this section,  we describe how the inclusive cross sections 
for the production 
of singlet states corresponding to  
the operators ${\cal O}^{I}$, 
through the scattering of particles in $\mathcal{N}=4$ SYM theory, can be used to obtain
various splitting functions and infrared safe coefficient functions.  
The generic scattering process in ${\cal N} =4$ SYM theory is given by
\begin{equation}
\label{eq:process}
a(p_1) + b(p_2) \rightarrow I(q) + \sum\limits_{i=1}^{m} X(l_{i}),
\end{equation}
where $a,b \in \{\lambda,g,\phi,\chi \}$ can be a Majorana or gauge or scalar or pseudoscalar particles. 
\begin{strikeout}
\st{$I$ represents a color singlet state denoted by half-BPS or Konishi or $T$ with invariant mass given by $Q^2=q^2$,
where $q$ is its four momentum.  } 
\end{strikeout}
$X$ denotes the final inclusive state comprising of 
$\{\lambda,g,\phi,\chi\}$. In the above equation, the momenta of the corresponding particles
are given inside their parenthesis  { with the invariant mass of the singlet state
denoted by $Q^2=q^2$. Except the singlet state all other particles are massless. }

The inclusive cross section, $\hat{\sigma}^I_{ab}(\hat s,Q^2,\epsilon)$, 
for the scattering process in Eq.~(\ref{eq:process}) in $4+\epsilon$ dimensions is given by
\begin{align}
\label{eq:sigma}
\hat{\sigma}^I_{ab}(\hat s,Q^2,\epsilon) =
 \frac{1}{2 \hat{s}} \int \left[ dPS_{m+1} \right]  \overline{\sum} \left|\mathcal{M}_{ab}\right|^2,
\end{align}
where $\hat{s} = (p_1+p_2)^2$ is the partonic center of mass energy. The phase space integration, $\int \left[dPS_{m+1}\right]$, is given by
\be
\label{eq:phasesp}
 \int \left[ dPS_{m+1} \right] = \int  \prod_{ i=1 }^{m+1} \frac{d^{n}l_i}{(2\pi)^{n}} 2\pi\delta_{+}(l_i^2-q_i^2)
(2\pi)^{n}
\delta^n
 \Big( \sum_{j=1}^{m+1} l_j - p_1 - p_2 \Big)\,, 
\ee
with $l_{m+1}=q$, $q_i^2 = 0$ for $i=1,\cdot \cdot \cdot m$ and $q_{m+1}^2=Q^2$.
The symbol $\overline{\sum}$ indicates sum of all the spin/polarization/generation 
and color of the final state particles $X$  and
the averaging over them for the initial state
scattering particles $a,b$. $\mathcal{M}_{ab}$ is the amplitude
for the scattering reaction depicted in Eq.~(\ref{eq:process}). 
We follow the Feynman diagrammatic approach to compute these amplitudes. 

The cross sections $\hat{\sigma}^I_{ab}$ can be expanded in powers of  
t'Hooft coupling constant $`a$' defined by 
\begin{eqnarray}
\label{coupl}
a  \equiv \frac{g^{2}N}{16\pi^{2}}\exp[\frac{\epsilon}{2}(\gamma_E-\ln4\pi)]  ,
\end{eqnarray}
where $N$ is the number of colors in SU(N) gauge theory and $\gamma_E=0.5772\cdot\cdot\cdot$, is the Euler-Mascheroni constant.
Note that the spherical factor that appears at every order in the perturbation theory resulting from the loop and 
phase space integrals, is absorbed into the coupling constant.
We compute the inclusive cross section order by order in perturbation theory as
\begin{eqnarray}
\hat \sigma^I_{ab}(z,Q^2,\epsilon) = \sum_{i=0}^\infty {a}^i  \hat \sigma^{I,(i)}_{ab}(z,Q^2,\epsilon) ,
\end{eqnarray}
where the scaling variable is defined by $z=Q^2/\hat s$.  
For the half-BPS and Konishi, at LO, only scalar and pseudoscalars contribute, 
but for the T, at LO,  all the particles namely Majoranas, gluons, scalars and 
pseudoscalars contribute.  At next-to-leading order (NLO) and next-to-next-to-leading order (NNLO)
level, there will be plethora of processes that will be available for study. 
At NLO, we need to evaluate the amplitudes involving purely virtual diagrams, called 
FFs and single real emission processes to the LO processes.  
For the NNLO, we need in addition the interference of processes with single real emission and one virtual 
loop with an emission.  
\begin{strikeout}
\st{Beyond LO, that is starting from order $a$, we encounter both virtual and real emission processes.
The virtual diagrams involve loop integrals and real emission ones, the phase space integrals.  }
\end{strikeout}
{Beyond LO, evaluation of the Feynman diagrams involves performing the loop integrals 
for the FFs and the phase space integrals arising in the real emission processes. }
Both the loop and  the phase space integrals are often divergent in four space-time dimensions due to the presence of 
UV and IR divergences, hence they need to be regulated.
Dimensional regularization (DR) has been quite successful
in regulating both UV as well as IR singularities, { where all the 
singularities  show up as poles in $\epsilon$. }
\begin{strikeout}
\st{If we regulate these integrals and express all the matrix elements in $d$ dimensions, the divergences show up as poles in $\epsilon$. } 
\end{strikeout}
There are several schemes of DR that exist.  In the scheme proposed by 
't Hooft and Veltman ~\cite{'tHooft:1972fi}, called DR scheme,
the gauge bosons in the loops are treated in $4+\epsilon$ dimensions with  $2+\epsilon$ helicity states
but the external physical ones in 4 dimensions having 2 helicity states.
In the  conventional DR scheme proposed by Ellis and Sexton ~\cite{Ellis:1985er}
one treats both the physical and unphysical gauge fields in  $4+\epsilon$ dimensions.
There is yet another scheme, namely the four dimensional helicity (FDH) scheme 
by ~\cite{Bern:1991aq,Bern:2002zk}  
wherein both the physical and unphysical gauge fields are treated in 4 dimensions.
In all these schemes the loop integrals are performed in
$4+\epsilon$ dimensions.  FDH scheme has been the most popular one in supersymmetric theories.

In this paper, we choose to work with the modified dimensional 
reduction ($\overline{\rm DR}$) scheme ~\cite{Siegel:1979wq, Capper:1979ns} which  protects the supersymmetry 
throughout.  In this scheme, the number of generations of scalar and pseudoscalar fields are such that
the resulting bosonic degrees of freedom is same as that of fermions, preserving the supersymmetry.
Since the gauge fields have $2+\epsilon$ degrees of freedom, there are $3-\epsilon/2$ scalars
and $3-\epsilon/2$ pseudoscalars in the regularised version of the theory so that the total
number of bosonic degrees of freedom in  ${d}$ dimensions is same as in four dimensions, 
namely 8.
It was shown in \cite{Ahmed:2016vgl} that this scheme has advantage  over the others
as it can be used even for operators that depend on space-time dimensions.  An example
of such an operator is the Konishi operator (see Eq.~(\ref{Op2})).
In \cite{Ahmed:2016vgl}, three-loop FFs of the Konishi operator was computed in
$\overline{\rm DR}$ scheme which correctly reproduces its anomalous dimensions up to the same level.  

In the $\overline{\rm DR}$ scheme,
in addition to analytically continuing the loop integrals of virtual amplitudes and phase space integrals
of real emission processes to $d$ space-time dimensions, 
all the traces of Dirac gamma matrices, flavour matrices $\alpha$ and $\beta$,
and various flavour sums/averages for the Majorana, scalar, pseudoscalar particles and 
polarisation sums/averages for the gauge fields are done in $d$ dimensions.  

The renormalisation of the fields and couplings are done with the help of renormalisation constants.  
 Due to supersymmetry, the coupling constant $g$  {does not require any renormalization, the beta function of the coupling is zero to all orders in the perturbation theory
\cite{Jones:1977zr,Poggio:1977ma}.}
Hence $\frac{{\hat a}}{\mu^{\epsilon}} = \frac{a}{\mu_R^{\epsilon}}$, 
where renormalization scale is denoted by $\mu_{R}$ and an arbitrary scale $\mu$ is introduced to keep the coupling dimensionless in $d$ dimensions.  
In addition, the amplitudes involving 
protected operators such as the half-BPS and the space-time conserved operator like T do not require overall renormalisation constant. 
Since the Konishi operator is not protected by supersymmetry,  we need to perform
an overall renormalisation order by order in perturbation theory.   
The corresponding renormalization constant $Z^{\mathcal K}\left( a(\mu_R), \epsilon \right)$,
satisfies the following renormalization group equation (RGE):
 %
\begin{align}
\label{eq:9}
&\frac{d\ln Z^{\mathcal K}}{d\ln\mu_R^2}  =
  \gamma^{\mathcal K} = \sum\limits_{i=1}^{\infty} a^{i}
  \gamma^{\mathcal K}_{i}\,.
\end{align}
The solution to the above equation is
\begin{equation}
\label{eq:17}
Z^{\mathcal K}  = \exp \Bigg(\sum_{n=1}^{\infty}a^n\frac{2\gamma^{\mathcal K}_n}{n\epsilon}\Bigg).
\end{equation}
Here $\gamma^K$ is the anomalous dimension whose value
up to two-loop was computed in ~\cite{Anselmi:1996mq, Eden:2000mv, Bianchi:2000hn} while the three-loop
results are available in~\cite{Kotikov:2004er, Eden:2004ua,Ahmed:2016vgl} 

The real emission processes start contributing from NLO, where any { one of the particles $\in$
$\{\lambda,g, \phi, \chi \}$}  can be emitted { ($m=1$ in Eq.~(\ref{eq:process})).} 
Note that at NNLO level, there will be two classes of real emission processes,
namely amplitudes with double real emissions ($m=2$ in Eq.~(\ref{eq:process}))
and those with {the interference of 
one real and one virtual associated with a radiation.}
The UV finite virtual amplitudes involving half-BPS, T and Konishi are sensitive to IR singularities.
The massless gluons can give soft singularities and 
the massless states in virtual loops  
can become parallel to one another, giving rise to collinear singularities.
\begin{strikeout}
\st{can have collinear configurations giving rise to collinear singularities.}
\st{The soft singularities from the virtual diagrams cancel against the those from the real emission processes, thanks to the Kinoshita-Lee-Nauenberg (KLN) theorem}~\cite{Kinoshita:1962ur, Lee:1964is}. 
\st{Similarly, the final state collinear singularities cancel among themselves in these 
inclusive cross sections leaving only initial state collinear singularities.}
\end{strikeout}
 { The soft and collinear singularities  from the virtual diagrams cancel against the soft and final 
state collinear divergences from the real emission processes, thanks to the KLN theorem}~\cite{Kinoshita:1962ur, Lee:1964is}.
Since the initial degenerate states are not summed in the scattering cross sections,
collinear divergences originating from incoming states remain as poles in $\epsilon$.  Hence,
like in QCD, the inclusive cross sections in ${\cal N}=4$ SYM theory, are singular in four dimensions. 
Following perturbative QCD~\cite{Collins:1985ue}, these singular cross sections
can be shown to factorize at the factorization scale $\mu_{F}$:
\begin{eqnarray}
\label{eq:massfact}
{\hat \Delta}^I_{ab}\left(z,Q^2,{1\over \epsilon}\right) &=&\left(\prod_{i=1}^3 \int_0^1 dx_i \right)
\delta\left(z-\prod_{i=1}^3 x_i\right)\,
~\sum_{c,d}\Gamma_{ca}\left(x_1,\mu_F^2,{1\over \epsilon}\right) 
\nonumber\\&&  
\times \Gamma_{db}\left(x_2,\mu_F^2,{1\over \epsilon}\right)
~  \Delta_{cd}^I\left(x_3,Q^2,\mu_{F}^2,\epsilon\right) ,
\end{eqnarray}
where the sum extends over the particle content $\{ \lambda, g, \phi, \chi \}$.
In the above expression $\hat{\Delta}^I_{ab}(z,Q^2,1/\epsilon) = \hat{\sigma}^{I}_{ab}(z,Q^2,\epsilon)/z$; 
the corresponding one after factorisation  
is denoted by $\Delta^I_{ab}$. 
If this is indeed the case, then we should be able to obtain $\Gamma_{ab}$ order by order
in perturbation theory from the collinear singular $\hat \Delta^I_{ab}$
by demanding $\Delta^I_{ab}$ is finite as $\epsilon \rightarrow 0$. 
The fact that the $\hat \Delta^I_{ab}$ are independent of the scale $\mu_F$ leads
the following RGE:
\begin{eqnarray}
\label{RGkernel}
\mu_F^2 {d \over d\mu_F^2}\Gamma(x,\mu_F^2,\ep)={1 \over 2}  P
                         \left(x\right) 
                        \otimes \Gamma \left(x,\mu_F^2,\ep\right),
\end{eqnarray}
where the function $P(x)$ is matrix valued and their elements $P_{ab}(x)$ 
are finite as $\epsilon \rightarrow 0$ and they are called splitting functions.   
This is similar to Dokshitzer-Gribov-Lipatov-Altarelli-Parisi (DGLAP) evolution equation~\cite{
Gribov:1972ri,Lipatov:1974qm,Altarelli:1977zs,Dokshitzer:1977sg, 
Floratos:1980hm, Floratos:1980hk, Curci:1980uw, Vogt:2004gi} in QCD for the parton distribution functions. 
In the  { $\overline{\rm DR}$} scheme, the solution to the RGE in terms of the splitting functions, {the latter expanded in $a$ as,}
\begin{equation}
\label{sfexpand}
 {P_{{ ca}}(x) = \sum_{i=1}^{\infty} a^i P^{(i-1)}_{{ ca}}(x)},
\end{equation}
can be found to  be
\begin{align}
\label{eq:kernel}
&\Gamma_{ca}\left(x, \mu_{F}^{2},{1 \over \epsilon}\right) = \sum\limits_{k=0}^{\infty}
  a^{k}\Gamma^{(k)}_{ca}\left(x, \mu_{F}^{2},{1 \over \epsilon}\right), \nonumber
\intertext{with}
&\Gamma_{ca}^{(0)} = \delta_{ca} \delta(1-x)\,,
\nonumber\\
&\Gamma_{ca}^{(1)} = \frac{1}{\epsilon}  P^{(0)}_{ca}(x)\,,
\nonumber\\
&\Gamma_{ca}^{(2)} = \frac{1}{\epsilon^{2}} \Bigg( \frac{1}{2} P^{(0)}_{ce} \otimes
  P^{(0)}_{ea} \Bigg) + \frac{1}{\epsilon}
  \Bigg( \frac{1}{2} P^{(1)}_{ca}\Bigg)\,. 
\end{align}
{Knowledge of $\hat \Delta_{cd}^I$ up to sufficient order both in $a$ as well as in $\epsilon$,
combined with the solution of Eq.~(\ref{RGkernel}) will give us the desired $P_{ab}^{(i)}(z)$, order by order 
in perturbation theory}.
Note that in the {$\overline{\rm DR}$} scheme, the AP kernels contain only ${1 \over \epsilon^n}$ where $n$ is positive definite.
\begin{strikeout}
\st{The splitting functions $P_{ab}^{(i)}(z)$ can extracted from
the collinear singular cross sections $\hat \Delta_{cd}^{I}(z,Q^2,1/\epsilon)$ by demanding
that $\Delta_{cd}^{I}(z,Q^2,\mu_F^2,\epsilon)$ is finite order by order in perturbation theory.}
\st{Hence, the splitting functions appearing in these kernels can be determined uniquely
provided $\hat \Delta_{cd}^I$ are known to sufficient order both in $a$ as well as in $\epsilon$. }
\end{strikeout}

In  Eq.~(\ref{sfexpand}) $c,a \in \{ \lambda, g, \phi, \chi \}$ thus
we have 16 splitting functions $P_{ab}$  at every order in perturbation theory.
To determine LO $P^{(0)}_{ab}$ and NLO
$P^{(1)}_{ab}$, we need to evaluate the scattering cross sections 
$\hat \sigma^I_{ab}$ for various choices of initial states $`ab$' up to second order
in the coupling constant $a$.  Since these are inclusive cross sections, sum over all the allowed final states need to
be done.  We find more than one splitting functions $P^{(i)}_{ab}$ appear
in single $\hat \Delta_{ab}^{I,(i)}$ which makes it difficult to determine them separately.  For example
the non-diagonal terms such as $P_{\lambda \phi}$ and $P_{\phi\lambda}$ would appear together
with some numerical coefficients in $\hat \Delta^{I,(i)}_{\lambda \phi}$, at every order.
We can disentangle  
them if we 
{ compute the contributions from more than one partonic cross sections,{ {\it i.e.} $I$ = half-BPS and T.} }
{In addition we have observed that $\hat \sigma^I_{\lambda \phi} = $ $\hat \sigma^I_{\lambda \chi}$,
 which is valid up to second order in $a$ for any $I$.}
 Hence, the number of $P_{ab}$ that we need to determine reduces to 10.  They
are given by $P_{gg}, P_{\lambda \lambda}, P_{\phi \phi}, P_{g \lambda}, P_{\lambda g}, P_{g \phi}, P_{\phi g}, P_{\lambda \phi},
P_{\phi \lambda}$ and $P_{\phi\chi}$.

The LO diagonal splitting functions $P_{cc}^{(0)}$ requires cross sections
$\hat \sigma_{cc}^{\text{T},(i)}$ with $i=0,1$ and the relevant processes are
\begin{eqnarray}
&&c+c \rightarrow \rm{T}, \quad \quad c+c \rightarrow \rm{T} + {\rm one\,\, loop},
\nonumber \\
&& c+c \rightarrow \text{T}+X,
\end{eqnarray}
where $X=g$ for $c\in\{\phi,g\}$ and $X\in \{g, \phi, \chi\}$ for $c=\lambda$.
Each { of the above} processes at $\mathcal{O}(a)$ contains only one $P_{cc}^{(0)}$, hence it is straightforward to 
obtain each of them independently.  
If we use the half-BPS operator, we can compute only $P_{\phi\phi}^{(0)}$ which we find
agrees with that obtained  { using the T operator.}
The non-diagonal LO splitting functions $P_{cb}^{(0)}$ requires the computation
of $\hat \sigma_{cb}^{\text{T},(i)}$ with $i=0,1$. At one loop the processes that contribute are given by
\begin{eqnarray}
c + b \rightarrow \text{T}+ c ,
\end{eqnarray}
where we have chosen:
${c \neq b}$ with $(c,b)\in\{(\lambda,\phi), (\lambda, g), (\phi, g)\}$. 
 {It is interesting to note that} in each { of the above}  subprocesses 
{ only the following combination of splitting functions appears:}
$\hat \sigma^{\text{T},(0)}_{cc} P^{(0)}_{cb} + \hat \sigma^{\text{T},(0)}_{bb} P^{(0)}_{bc}$.  We can
disentangle $P^{(0)}_{cb}$ and $P^{(0)}_{bc}$ separately by comparing the coefficients  of $\hat \sigma^{\text{T},(0)}_{cc}$  
and $\hat \sigma^{\text{T},(0)}_{bb}$. The remaining LO splitting function $P^{(0)}_{\phi\chi}=P^{(0)}_{\chi \phi}
$ is found to be identically zero as they start at  ${\cal O}(a^2)$. 

At NLO level, the diagonal splitting function $P^{(1)}_{cc}$ requires the computation of
$\hat \sigma^{\text{T},(2)}_{cc}$ and 
$\hat \sigma^{\text{T},(i)}_{cb}$ with $i=0,1$,
for different combinations of $c$ and $b$.  $\hat \sigma^{\text{T},(2)}_{cc}$ gets contribution from
two-loop virtual processes
\begin{eqnarray}
c + c \rightarrow \text{T} + {\rm two\, loops}\,,
\end{eqnarray}
one-loop with a single real emission processes
\begin{eqnarray}
c + c \rightarrow \text{T} + X + {\rm one \, loop},
\end{eqnarray}
where $X=g$ for $c\in\{\phi,g\}$, $X\in \{g, \phi, \chi\}$ for $c=\lambda$ and pure double emission processes
\begin{eqnarray}
c+ c \rightarrow \text{T} + b + b   ,
\end{eqnarray}
where for every pair of initial states {made up of a pair of $c$'s with}
$c = \lambda,g,\phi$, 
the allowed final states contain a pair of $b$'s where $b = \lambda,g,\phi$.    
Since the half-BPS operator couples to only $\phi$'s at LO,  we can compute $P^{(1)}_{\phi \phi}$
from $\hat \sigma^{\rm{BPS},(2)}_{\phi\phi}$ as well.  This provides an independent check on our
results.  

Unlike the diagonal splitting functions, the non-diagonal ones can not be determined from $\hat \sigma^{\text{T},(2)}_{cb}$
alone.  The cross sections $\hat \sigma^{\text{T},(2)}_{cb}$ where $c\not=b$   
always contain the combinations of $P^{(1)}_{cb}$ and $P^{(1)}_{bc}$.  Hence determining them from
single cross section is not  possible.  Therefore we resort to    
$\hat \sigma^{\rm{BPS},(2)}_{cb}$ which can give $P^{(1)}_{cb}$ unambiguously.  Knowing $P^{(1)}_{cb}$ and
using $\hat \sigma^{\text{T},(2)}_{cb}$, we determine $P^{(1)}_{bc}$.  The relevant processes to determine 
$P^{(1)}_{c \lambda}$ and $P^{(1)}_{\lambda c}$ where $c=g,\phi$  are given by
\begin{eqnarray}
&&\lambda + c  \rightarrow I +  \lambda + {\rm one\, loop} ,
\nonumber\\
&&\lambda + c \rightarrow I +  \lambda + b ,
\end{eqnarray} 
where $b\in\{\phi, \chi, g\}$ and $I=$ T, half-BPS.
The cross sections, $\hat \sigma^{I}_{g \phi}$ where $I=$T, half-BPS that contribute to $P^{(1)}_{\phi g}$ and $P^{(1)}_{g \phi}$ can be obtained  
.  The relevant processes are 
\begin{eqnarray}
&&g +\phi \rightarrow I+ \phi + {\rm one\, loop},
\nonumber\\
&&g+\phi \rightarrow I+ g + \phi ,
\nonumber\\
&&g+\phi \rightarrow I + \lambda +\lambda .
\end{eqnarray}
Finally, the splitting function $P^{(1)}_{\phi \chi}=P^{(1)}_{\chi \phi}$ is obtained from
the cross sections $\hat \sigma^{I}_{\phi \chi}$  with $I=$T, half-BPS
which get contributions from the subprocesses
\begin{eqnarray}
&&\phi +\chi \rightarrow I + \phi +\chi ,
\nonumber\\
&&\phi +\chi \rightarrow I + \lambda +\lambda. 
\end{eqnarray}
In QCD, the kernel $\Gamma_{ab}$ contains 9 different splitting functions
because $a,b\in\{q, \overline q, g\}$ for a given flavour quark.  The Mellin moments of them
namely 
\begin{eqnarray}
\label{spano}
\int_0^1 dz z^{j-1}P_{ab}(z) = \gamma_{ab,j},
\end{eqnarray}     
are anomalous dimensions of gauge invariant local operators made up of quark, anti-quark and gluon
fields, see \cite{Politzer:1974fr, Buras:1979yt,Altarelli:1981ax,Hagiwara:1984jk,Georgi:1951sr,Gross:1974cs}.  
Following QCD, we can relate the Mellin moments of $P_{ab}$ obtained
in ${\cal N}=4$ SYM theory with the anomalous dimensions of composite operators given by  
\begin{eqnarray}
{\cal O}^\lambda_{\mu_1\cdot \cdot \cdot \mu_j} &=&
S\left\{\overline \lambda^a_m \gamma_{\mu_1} D_{\mu_2} \cdot \cdot \cdot D_{\mu_j} \lambda^a_m\right\}\,,
\\
{\cal O}^g_{\mu_1\cdot \cdot \cdot \mu_j}&=&
S\left\{G^a_{\mu \mu_1}  D_{\mu_2} \cdot \cdot \cdot D_{\mu_{j-1}} G^{a\mu}_{\mu_j}\right\} \,,
\\
{\cal O}^{\phi}_{\mu_1\cdot \cdot \cdot \mu_j}&=&
S\left\{\phi^{a}_i D_{\mu_1} \cdot \cdot \cdot D_{\mu_j} \phi^a_i \right\}\,,
\\
{\cal O}^{\chi}_{\mu_1\cdot \cdot \cdot \mu_j}&=&
S\left\{\chi^a_i D_{\mu_1} \cdot \cdot \cdot D_{\mu_j} \chi^a_i\right\} \,.
\end{eqnarray}
The symbol $S$ indicates symmetrisation of indices $\mu_1 \cdot \cdot \cdot \mu_j$.
Note that these operators mix under renormalisation and the corresponding anomalous dimensions are
given by $\gamma_{ab,j}$.
In addition, when $j=2$, the sum reproduces the gauge invariant part of energy momentum tensor  
which does not require any overall renormalisation.   In other words, the sum 
$\sum_a {\cal O}^a_{\mu_1\mu_2}$ is UV finite, hence   
\begin{eqnarray}
\mu_R^2 {d \over d\mu_R^2} \left(\sum_a {\cal O}^a_{\mu_1\mu_2}\right) = 0, \quad \quad a \in \{\lambda,g,\phi,\chi\}.
\end{eqnarray}
This implies
\begin{eqnarray}
\label{anomdim}
\sum_a \gamma_{ab,2} = 0 \quad \quad \quad a,b \in \{\lambda,g,\phi,\chi\}.
\end{eqnarray}
We will show that splitting functions computed in the present paper 
satisfy the above relation up to NLO level, 
namely at each perturbative order  $i$
\begin{eqnarray}
\label{sfmellin}
\sum_{a} \int_0^1 dz\, z P^{(i)}_{ab}(z)  = 0\,,  \,\, \text{where} \quad i=0,1,
\end{eqnarray}
with $a,b$ given in Eq.~(\ref{anomdim}). In the next section, we shall discuss the methodology that we have adopted 
to compute the individual partonic cross sections $\hat{\sigma}^{I}_{bc}$.
  
\section {Methodology}
\label{method}
The computation of $\hat{\Delta}^I_{ab}(z,Q^2,\epsilon)$ {\it i.e.} ${\hat \sigma}^I_{ab}(z,Q^2,\epsilon)/z$ 
beyond the LO involves evaluating processes with real emissions 
and virtual loops. 
We generate relevant  Feynman diagrams by using the package QGRAF~\cite{Nogueira:1991ex}.
The raw output from QGRAF is converted to a suitable format for further manipulation
by using our in-house codes written in FORM ~\cite{Vermaseren:2000nd, 
Tentyukov:2007mu}. We then compute the square of the diagrams by summing over the spins of Majoranas,   
polarization vectors of gluons and generation indices of Majoranas, scalars and pseudoscalars. 
In addition, we sum the colors of all the external states.  
The resulting expression contains large number of 
Feynman integrals and phase space integrals.   Using a Mathematica based package  
LiteRed~\cite{Lee:2012cn, Lee:2013mka}
we reduce all the Feynman integrals to few Master Integrals (MIs). 
While there were brisk developments in evaluating the 
loop diagrams, progress in computing the phase space integrals for real emission processes took place slowly.
It is worthwhile to mention that the NNLO QCD corrections to DY pair production 
\cite{Hamberg:1990np} was achieved by 
choosing Lorentz frames in such a way that the integrals can be achieved. 
An alternate approach was proposed in ~\cite{Harlander:2002wh} to obtain 
the inclusive production of Higgs boson.  
In this approach, the phase space integrals were done after expanding the matrix elements around the scaling variable
$z=1$.  These approaches pose the problem of dealing with large number of integrals.  
An elegant formalism was developed by Anastasiou
and Melnikov \cite{Anastasiou:2002yz} which helps to reduce these large number of 
phase space integrals to a set of few master integrals.   In this formalism, the phase space integrals are 
first converted to loop integrals by employing 
the method of reverse unitarity. 
One replaces the $\delta_{+}$ functions,
coming from phase space integrals (see Eq.~(\ref{eq:phasesp})), by the difference of  propagators,
\begin{align}
\label{eq:revunita}
\delta_{+}(q^{2}-m^{2}) \sim \frac{1}{q^{2}-m^{2}+i\varepsilon} -
  \frac{1}{q^{2}-m^{2}-i\varepsilon} .
\end{align}
This replacement results in loop integrals which    
can be simplified to fewer set of MIs with the help of integration by parts (IBP) identities. 
Care is needed while using IBP identities because the shifts of momenta are not 
allowed.  The MIs that remain after the application of IBP identities
are transformed back to phase space integrals by
appropriately replacing those propagators that were introduced in place of $\delta_+$ functions.  
Since the number of integrals at this stage is much smaller, the problem reduces to evaluation of
fewer integrals using standard techniques.   
The phase space integrals relevant up to NNLO level can be found in~\cite{Pak:2011hs}.
We used this approach to obtain $\hat{\sigma}^I_{ab}$
up to ${\cal O}(a^2)$ in perturbation theory.  For more details on the implementation, see 
\cite{Anastasiou:2002yz,Ahmed:2016qhu,Banerjee:2017ewt}. 
\section {Analytical results and discussion}
\label{results}
The splitting functions $P_{ab}^{(i)}(z)$ for $i=0,1$ are extracted from
the collinear singular cross sections $\hat \Delta_{cd}^{I}(z,Q^2,1/\epsilon)$ by demanding
that $\Delta_{cd}^{I}(z,Q^2,\mu_F^2,\epsilon)$ are finite order by order 
in perturbation theory.  In the  $\overline{\rm DR}$ scheme,
at LO level, the diagonal ones are found to be
\begin{eqnarray}
\label{eq:expresf1l}
P_{\lambda \lambda}^{\,(0)}(z) &=& 8 \left[\T(z) + 3 -2z\right],
\nonumber\\
P_{gg}^{\,(0)}(z) &=& 8  \left[1 - \V(z) + \T(z) + z (1-z) \right], 
\nonumber\\
P_{\phi \phi}^{\,(0)}(z) &=& 8  \left[\T(z) +1\right],
\end{eqnarray}
and the non-diagonal ones are
\begin{align}
P_{g \lambda}^{\,(0)}(z) &= 4  \left[ z - 2 \V (z) \right], 
\quad\quad
P_{\lambda g}^{\,(0)}(z) = 16  \left[ 1 - 2z(1-z) \right],
\nonumber\\
P_{\phi \lambda}^{\,(0)}(z) &= 6  z,
\quad\quad\quad\quad\quad\quad\,\,\,\,
P_{\lambda \phi}^{\,(0)}(z) = 16 , 
\nonumber\\
P_{\phi g}^{\,(0)}(z) &= 12  z \left(  1 - z \right), 
\quad \quad\,\,\,\,\,
P_{g\phi}^{\,(0)}(z)  = - 8  \, \V(z).
\end{align}
The LO splitting functions involving $\chi$ are obtained using
\begin{align}
P_{\chi \chi}^{\,(0)}(z) &= P_{\phi \phi}^{\,(0)}(z)\, , 
\quad\quad
P_{\chi \phi}^{\,(0)}(z)  = P_{\phi \chi}^{\,(0)}(z) =0,
\nonumber\\
P_{b \chi}^{\,(0)}(z) &= P_{b \phi}^{\,(0)}(z)\,,
\quad\quad
P_{\chi b}^{\,(0)}(z) = P_{\phi b}^{\,(0)}(z)\,\text{where}\,\,b \in \{\lambda,g\}.
\end{align}
%
The extraction of the splitting functions at NNLO level involves use of both the half-BPS as well as T operators
because of the presence of more than one splitting functions in a single cross section.  
By appropriately choosing the singlet final states and 
the corresponding pair of particles in the initial states, as described in the previous section, we obtain
\begin{eqnarray}
\label{eq:expresf2l}
P_{\lambda \lambda}^{\,(1)}(z)  &=&  
24\,  \zeta_{3}\, 
\delta(1-z) + 8 \left[\log^2(z)  - 2\, \zeta_2 \right]\left[\T(z) + \T(-z) +6 \right]- 32\, \log(z) \log(1- z)
\nn  
&&\times\left[\T(z) + 3 - 2z \right]- 32 \left[\text{Li}_2(-z) + \log(z) \log(1+ z)\right] \left[\T(-z) 
+ 3 + 2z \right]   
\nn 
&&+ 64 \log(z) \left[3 + z + {4\over 3} z^2 \right] + {640\over 9}\frac{1}{z}  + 128 z 
- {1792 \over 9} z^2 
,
\nn 
P_{g \lambda}^{\,(1)}(z) &=& \,   
32\zeta_2 + 16  \left[\text{Li}_2(-z) + \log(z) \log(1+ z)\right]
\left[2 \V(-z) + z \right]-16 \log^2(z)
\nn  
&&+  16 \, \log(z) \log(1-z) \left[2 \V(z) - z \right]
 - 16 \log(z) \left[9 + 2 z + {4 \over 3}z^2 \right] 
 \nn  
&& + 80 - {1072 \over 9}\frac{1}{z} + {352\over 9} z^2  
,
 \nn 
 P_{\phi \lambda}^{\,(1)}(z) &=& \,   
24z \left[ \text{Li}_2(-z) + \log(z) \log(1+ z) -  \log(z) \log(1-z) \right] -  8 \log(z) \left[3 + 2z+ 4 z^2 \right] 
 \nn 
&&  + 16 + 24 \V(-z) - 64 z  + 80 z^2,
 \nn 
P_{gg}^{\,(1)}(z) &=&  \,  
24 \zeta_3 \delta(1-z) +\left[2\zeta_2 -  \log^2(z)\right] \left[64 - 8 \T(-z) - 8 \T(z) +16z^2  \right] 
\nn 
&&+ 32 \left[\text{Li}_2(-z)+ \log(z) \log(1+ z) \right] \left[\V(- z) - \T(-z)- 1 + z + z^2 \right]  
\nn 
&&+ 32 \left[\log(z)\log(1- z)\right] \left[\V(z) -\T(z )-1 - z + z^2 \right]
\nn
&&- \log(z) \left[144 + 112 z 
- {352 \over 3} z^2 \right]+ 80 -{1072 \over 9}\frac{1}{z} 
- 208z + {2224 \over 9} z^2  
,
\nn
P_{\lambda g}^{\,(1)}(z) & = & \,  
\left[\log^2(z)  - 2\zeta_2 \right] \left[32 + 64z^2 \right]
-   64\left[\text{Li}_2(-z) + \log(z) \log(1+ z)\right] \left[1+2z+2z^2  \right]
\nn 
&&
 - 64\left[\log(z) \log(1- z) \right] \left[1 - 2z +2z^2 \right] 
 +  \log(z) \left[192 + 320 z + {1792 \over 3}z^2 \right]
 \nn 
&&  + {640\over 9}\frac{1}{z} + 896z - {8704 \over 9}z^2 
,
 \nn 
 P_{\phi g}^{\,(1)}(z) & = & \,  
24 z^2 \left[2 \zeta_2 - \log^2(z) \right]  + 
48 z (1+z)\left[\text{Li}_2(-z) + \log(z) \log(1+ z)\right]
 \nn 
&& - 48 z(1 - z) \log(z) \log(1- z)
 -  \log(z) \left[24 + 104 z + 240 z^2\right]  
 \nn 
&& -64+ 24 \V(-z) -344z +360z^2 
,  
 \nn 
 P_{\phi \phi}^{\,(1)}(z) & = & \,  
24 \zeta_3 \delta(1-z) + \left[ 8 \log^2(z) - 16 \zeta_2 \right]
 \left[\T(z) + \T(-z) + 2 \right]  
 \nn 
&& -  32 \left[\text{Li}_2(-z)+ \log(z) \log(1+ z)\right] \left[\T(-z) 
  +1\right] -32 \left[\T(z)+1\right] \log(z) \log(1- z)
  \nn 
&&  + \log(z) 
  \left[-24 + 24z + 16z^2 \right] - 64  + 24 \V(-z) + 40 z -24 z^2  
,
 \nn 
 P_{\lambda \phi}^{\,(1)}(z) &= & \,  
32 \left[ \log^2(z) - 2 \zeta_2 - 2 \text{Li}_2(-z) - 2  \log(z) \log(1+ z)    
  - 2  \log(z) \log(1- z)  \right]
  \nn 
&&  + \log(z) \left[192 - 64z 
  -{128 \over 3} z^2  \right] + {640 \over 9}\frac{1}{z} -128z + {512 \over 9} z^2 
,
 \nn 
 P_{g  \phi}^{\,(1)}(z) &= & \,  
16 \left[2\zeta_2 -  \log^2(z) \right] +
 32\left[\text{Li}_2(-z) + \log(z) \log(1+ z)\right] \V(-z) 
 \nn 
&& + 32\V(z)\log(z)\log(1- z)
 + \log(z) \left[ -144 + 16 z + {32\over 3} z^2 \right]
\nn 
&&+ 80 -{1072 \over 9}\frac{1}{z} + 48 z -{80\over 9} z^2
,
 \nn 
  P_{\phi\chi}^{\,(1)}(z) &=& \,  
8\log(z) \left[-3 + 3 z + 2z^2 \right] + 24\,\V(-z) - 64 + 
40z -24z^2\,,
\end{eqnarray}
and the splitting functions involving $\chi$ are obtained using
\begin{align}
P_{\chi \chi}^{\,(1)}(z) &= P_{\phi \phi}^{\,(1)}(z)\, , 
\quad\quad
P_{\chi \phi}^{\,(1)}(z)  = P_{\phi \chi}^{\,(1)}(z),
\nonumber\\
P_{b \chi}^{\,(1)}(z) &= P_{b \phi}^{\,(1)}(z)\,,
\quad\quad
P_{\chi b}^{\,(1)}(z) = P_{\phi b}^{\,(1)}(z)\,\text{where}\,\,b \in \{\lambda,g\}.
\end{align}
%
In above $\T(z) = 1/(1-z)_+ -2$ and $\V(z) = 1-1/z$. The action of ``+ distribution" on a dummy function $f(z)$ is defined by
\begin{eqnarray}
\label{plusdist}
\int_0^1 dz f(z) \left[{\log^n(1-z) \over 1-z}\right]_+
=\int_0^1 dz \left[f(z)-f(1)\right] {\log^n(1-z) \over 1-z}.
\end{eqnarray}
We find that the both LO and NLO splitting functions satisfy the following relations: 
\begin{eqnarray}
\sum_{a=\lambda,g,\phi,\chi} P^{(i)}_{a \lambda} = 
\sum_{a=\lambda,g,\phi,\chi} P^{(i)}_{a g} = 
\sum_{a=\lambda,g,\phi,\chi} P^{(i)}_{a\phi} = 
\sum_{a=\lambda,g,\phi,\chi} P^{(i)}_{a\chi}  
=I^{(i)}(z) ,
\quad \quad i=0,1,
\end{eqnarray}
where 
\begin{eqnarray}
I^{(0)}(z) &=& 8  \Bigg[ {1 \over (1-z)_+} + \frac{1}{z}\Bigg],
\nonumber\\
I^{(1)}(z) &=&   
24\zeta_3\delta(1-z) + 32 {1 \over z}\left[\text{Li}_2(-z) + \log(z) \log(1+z) - \log(z) \log(1-z)\right]
\nonumber\\
&&+{1 \over (1-z)_+} \left[-32 \log(z) \log(1-z) + 8 \log^2(z) - 16 \zeta_2\right]
\nonumber\\
&&+{1 \over 1+z} \left[-32 \text{Li}_2(-z) -32 \log(z) \log(1+z) + 8 \log^2(z) -16 \zeta_2\right].
\end{eqnarray}
Using the above relations, we confirm the identity given in Eq.~(\ref{sfmellin}) {\it i.e.}
\begin{eqnarray}
\label{check2}
\sum_{a=\lambda,g,\phi,\chi} \int_0^1 dz \, z P^{(i)}_{ab} = \int_0^1 dz \, z I^{(i)}(z) = 0\,, \quad i=0,1\,\,\text{and}\,\, b = \{\lambda,g,\phi,\chi\}.
\end{eqnarray}
We find that  both at NLO and NNLO, only the diagonal splitting functions contain ``+'' distributions.  In addition,
at NNLO level, terms proportional to $\delta(1-z)$ start contributing to diagonal splitting functions.  Hence,
 {in the limit z $\rightarrow$ 1}, the diagonal splitting functions can be parametrized as
\begin{eqnarray}
\label{diagsf}
P^{(i)}_{aa}(z) = 2 A_{i+1} {1 \over (1-z)_+} + 2 B_{i+1} \delta(1-z) + R^{(i)}_{aa}(z),
\end{eqnarray}
where  $A_{i+1}$ and $B_{i+1}$ are the cusp 
\cite{Korchemsky:1987wg,Beisert:2006ez,Correa:2012nk,Ahmed:2016vgl}
 and collinear \cite{Ahmed:2016vgl} anomalous dimensions respectively. $R^{(i)}_{aa}(z)$ is the regular 
function as $z\rightarrow 1$.  We find that 
\begin{eqnarray}
A_1 &=& 4, A_2 = -8 \zeta_2 \,, \quad \quad \text{and} \quad \quad
B_1 = 0, B_2 = 12 \zeta_3\,,
\end{eqnarray} 
which are in agreement with the result obtained from the FFs of the half-BPS operator
\cite{Korchemsky:1987wg,Beisert:2006ez,Correa:2012nk, Ahmed:2016vgl}.  

Using the supersymmetric extensions of Balitskii-Fadin-Kuraev-Lipatov (BFKL) \cite{Lipatov:1976zz, Fadin:1975cb, Balitsky:1978ic}
and DGLAP  \cite{Gribov:1972ri,Lipatov:1974qm,Altarelli:1977zs,Dokshitzer:1977sg}
evolution equations,
Kotikov and Lipatov ~\cite{Kotikov:2000pm,Kotikov:2002ab,Kotikov:2004er,Kotikov:2003fb,
Kotikov:2006ts} conjectured leading transcendentality (LT) principle 
which states that the eigenvalues of anomalous
dimension~\cite{Marboe:2016igj} matrix of twist two composite operators made out of $\lambda$, $g$ and complex $\phi$ fields 
in ${\cal N}=4$ SYM theory contain uniform transcendental terms at every order in perturbation theory. 
{Interestingly }they are related to 
 { the corresponding quantities in} QCD \cite{Moch:2004pa,Vogt:2004mw}. In~\cite{Kotikov:2004er} it has been shown that the eigenvalues of the anomalous dimension matrix are related to the universal anomalous dimension by shifts in spin-$j$ up to three-loop level. Unlike~\cite{Kotikov:2003fb}, we distinguish scalar and pseudo-scalar fields and compute their anomalous dimensions and their mixing in Mellin-$j$ space. We find two of the eigenvalues of the resulting anomalous dimension matrix coincide with the universal eigenvalues obtained in~\cite{Kotikov:2003fb} after finite shifts and the remaining two coincide with the universal ones only in the large $j$ limit (i.e. $z\rightarrow 1$). For reference, we explicitly present the eigenvalues computed in this paper in appendix~\ref{appB}.
One can associate the transcendentality weight $n$ to terms such as $\zeta(n)$, $\epsilon^{-n}$ and
also to the weight of the harmonic polylogarithms  that appear in the perturbative calculations.
Similar relations were found in certain scattering amplitudes \cite{Bern:2006ew,Naculich:2008ys}, FFs of BPS type operators \cite{Bork:2010wf,Gehrmann:2011xn,Brandhuber:2012vm,Eden:2012rr},
light-like Wilson loops \cite{Drummond:2007cf,Drummond:2013nda}
and correlation functions \cite{Eden:2012rr,Drummond:2013nda,Basso:2015eqa,Goncalves:2016vir,Basso:2017muf,Georgoudis:2017meq} 
computed in ${\cal N}=4$ SYM theory.  
It is shown that in \cite{Brandhuber:2012vm}, the two-loop three-point MHV FFs of the half-BPS operator 
have uniform transcendental terms in the finite reminder functions.  
Several FFs in QCD when $C_A=C_F=n_f=N$ coincide with certain FFs in ${\cal N}=4$ SYM theory,
and the LT terms of
the amplitude for Higgs boson decaying to three on-shell gluons in
QCD ~\cite{Gehrmann:2011aa,Koukoutsakis:2003nba} are related to
the two-loop three-point MHV FFs of the half-BPS operator~\cite{Brandhuber:2012vm}. Similar correspondence was shown between two-loop three-point FF~\cite{Banerjee:2016kri} of the half-BPS operator and the pseudoscalar Higgs boson plus three-gluon amplitudes~\cite{Banerjee:2017faz} in minimal supersymmetric SM.
Two-point FFs of quark current operator~\cite{Gehrmann:2010ue}, pseudoscalar~\cite{Ahmed:2015qpa} operators, energy momentum tensor~\cite{Ahmed:2015qia,Ahmed:2016qjf} of the QCD up to three loops also show the
same behaviour.  It was shown in \cite{Ahmed:2016vgl,Banerjee:2016kri}, unlike BPS operators, 
the Konishi operators do not have uniform transcendental terms but their LT terms in FFs between
$\phi \phi$ and in remainder function computed between $g\phi\phi$ external state coincide with
the corresponding ones of the half-BPS. 

As can be seen from the results of splitting functions (see Eq.~(\ref{eq:expresf2l})), 
at each order $n$, the splitting functions consist of 
terms which have trancendentality ranging  from $2 n $ to 0. 
It is worth comparing the splitting functions in ${\cal N}=4$\ SYM theory, $P_{ab}$ 
with the ones obtained 
in QCD, $P_{ab}^{\rm{QCD}}$. 
We apply the following  color transformation on the QCD ones for comparison: $C_{A}=C_{F} =n_f= N$. We find that the one loop splitting functions $P_{gq}^{\rm{QCD},(0)}$ and  $P_{gl}^{(0)}$ 
are identical;
$P_{qg}^{\rm{QCD},(0)}$ and  $P_{lg}^{(0)}$ are also identical up to an overall factor.
For $P_{qg}^{\rm{QCD},(1)}$ and $P_{lg}^{(1)}$, apart from an overall factor, 
we find that only terms proportional to $ \log^2(z)$ are  different.
We also observe that LT parts of   $P_{gq}^{\rm{QCD},(1)}$ and $P_{gl}^{(1)}$ 
differ only in their $ \log^2(z)$ terms.

We now move on to study the finite cross sections $\Delta^I_{ab}$ up to NNLO level.
These cross sections
are computed in power series of the coupling constant $a$ as
{
\begin{equation}
\Delta^I_{ab} = \delta(1-z)\delta_{ab} + a~\Delta^{I,(1)}_{ab} + a^2~\Delta^{I,(2)}_{ab} + \cdot \cdot \cdot\\
\end{equation}
These $\Delta_{ab}^{I,(i)}$ contain both regular functions as well as 
distributions in the scaling variable $z$.  
The former are made up of polynomials and multiple polylogarithms of $z$ that are
finite as $z\rightarrow 1$ and they are from hard particles.
The distributions are from soft and collinear particles, which
show up at every order in the perturbation theory in the form of $\delta(1-z)$ and ${\cal D}_i(z)$ 
where
\begin{eqnarray}
{\cal D}_i(z) = \left( {\log^i(1-z) \over 1-z}\right)_+,
\end{eqnarray}
 {and its action on a regular function is shown
 in Eq.~(\ref{plusdist}).} { More precisely these distributions 
 originate from the real emission processes through
 
 \begin{equation}
\label{matrixplus}
(1-z)^{-1+\epsilon} = \frac{1}{\epsilon} \delta(1-z) + 
\sum_{k=0}^{\infty} \frac{\epsilon^{k}}{k!} {\cal D}_{k}.
\end{equation}
These distributions constitute what is called the 
threshold or soft plus virtual (SV) part of the cross section, denoted by 
$\Delta_{ab}^{\text{SV}}$.
We can  now express the total cross section as,
\begin{eqnarray}
\Delta_{ab}^{I,(i)} = \Delta^{I,(i), \text{SV}}_{ab} + \Delta^{I,(i), \rm{Reg}}_{ab},
\end{eqnarray}
where
\begin{equation}
\label{deltaSV}
\Delta^{I,(i),\text{SV}}_{ab}= 
\delta_{ab} \left (c^{I}_{i}\delta(1-z)+ \sum_{j=0}^{2i-1} d^I_{ij}\mathcal{D}_{j}(z) \right).
\end{equation}
The constants $c^{I}_{i}$ and $d^I_{ij}$ are absent when $a\not = b$.  For the diagonal ones $(a=b)$, they 
depend on the final singlet state $I$ and are in general
functions of rational terms and irrational $\zeta$. 
For the diagonal ones, $\Delta_{aa}^{I,(i),\rm{SV}}$ are found identical to
each other for $I=$\,BPS, T.  Up to NNLO level, they are found to be 
\begin{eqnarray}
\Delta_{aa}^{I,(0),\rm{SV}} &=&\delta(1-z)\,, \nonumber \\
\Delta_{aa}^{I,(1),\rm{SV}} &=&8\zeta_2\delta(1-z) + 16{\cal D}_1(z)\,, \nonumber \\
\Delta_{aa}^{I,(2),\rm{SV}} &=& 
-{4 \over 5}\zeta_{2}^{2}\delta(1-z)
+312\zeta_3{\cal D}_0(z) 
- 160\zeta_2 {\cal D}_1(z)
+ 128{\cal D}_3(z).
\end{eqnarray}
We observe that at every order, the above terms demonstrate uniform transcendentality which is
{ 1 at NLO and  3 at NNLO. Note that $\delta(1-z)$ has -1 transcendental weight
which can be understood 
from Eq.~(\ref{matrixplus}) by noting that the term $\epsilon^{-n}$ has transcendentality
$n$. We also notice that the highest distribution at every order determines the transcendental weight
at that order.}
It is interesting to note that the above coefficient functions are exactly identical to the
LT parts of the corresponding result in the SM for the Higgs boson production through gluon fusion computed in the effective
theory, upon proper replacement of the color factors in the following way {\it i.e.} $C_A=C_f=n_f=N$.
On the other hand for $I=\mathcal{K}$, we find
up to NNLO level, 
\begin{eqnarray}
\Delta_{aa}^{\mathcal{K},(0),\rm{SV}} &=&\delta(1-z)\,, \nonumber \\
\Delta_{aa}^{\mathcal{K},(1),\rm{SV}} &=&\left[-28 + 8\zeta_2\right]\delta(1-z) + 16{\cal D}_1(z)\,, \nonumber \\
\Delta_{aa}^{\mathcal{K},(2),\rm{SV}} &=& 
\left[604 -272 \zeta_2-{4 \over 5}\zeta_{2}^{2}\right]\delta(1-z)
+312\zeta_3{\cal D}_0(z) 
\nonumber\\
&&-\left[160\zeta_2 +448\right] {\cal D}_1(z)
+ 128 {\cal D}_3(z).
\end{eqnarray} 
Unlike BPS and T type, for Konishi, $\Delta^{\mathcal{K},{(i)},\rm SV}_{aa}$  does not have uniform transcendentality but
its LT terms coincide with those of BPS/T. 

\begin{strikeout}
\st{In perturbative QCD, the fixed order predictions for the observables 
are often unreliable in certain regions of phase space due to 
the presence of large logarithms }\cite{arXiv:1805.01186v1}. 
\st{For example, at the hadron colliders,
the inclusive observables like total cross section or invariant mass distribution of finial state
colorless state and some differential distributions contain large logarithms which can spoil
the reliability of fixed order results.}
\st{For example, at the partonic threshold i.e. 
when the initial partons have just enough energy to 
produce the final state colorless particle and soft gluons, 
the phase space available for the gluons become severely constrained 
giving large logarithms.  In the resummation approaches 
these large logarithms are systematically 
resummed to all orders in perturbation theory leading to reliable predictions. }
\end{strikeout}
The SV part of the inclusive observables in QCD is well understood to all orders in perturbation
 theory.  For example, the SV part of the inclusive cross section gets contribution from virtual part, namely 
the form factor and the soft, collinear configurations of the real emission processes.
In these observables, the soft singularities cancel between virtual and real emission processes, 
while the initial  collinear ones are removed by
 mass factorisation, thus giving IR finite results. Interestingly,  
 the factorisation property 
 of these cross sections can be used to identify the process independent soft distribution function 
 which depends only the incoming states.  In addition, they satisfy certain differential equation similar
to K+G equation of FFs.  
The solution gives all order prediction for the soft part of the observable 
in terms of soft anomalous dimensions $f_a$ with $a=q,g$.
Following \cite{Ravindran:2005vv} and noting that only $\Delta^{I, \rm SV}_{aa}$ contains threshold logarithms,
its all order structure can be expressed as
\begin{eqnarray}
\label{SVravi}
\Delta^{I, \text{SV}}_{aa} &=&  \left(Z^{I}\left(a,\epsilon\right)\right)^2 |\hat{F}^{I}_{aa}(Q^2,\epsilon)|^2\delta(1-z)\otimes {\cal C} \exp \left(2\Phi^I_{aa}(z,Q^2,\epsilon)\right)
\nonumber \\
&& \otimes\Gamma^{-1}_{aa}(z,\mu_F^2,\epsilon)\otimes \Gamma^{-1}_{aa}(z,\mu_F^2,\epsilon).
\end{eqnarray}
In above $I$ can be any one of the three operators considered in our current work. 
$Z^{I}(a,\epsilon)$ is the overall operator renormalization constant, which is unity
for $I=$ half-BPS and T operators; however, for $I={\mathcal K}$, up to three loop, the pertubative coefficients of 
$Z^{{\cal K}}$ are available ~\cite{Anselmi:1996mq, Eden:2000mv, Bianchi:2000hn,Kotikov:2004er, Eden:2004ua,Ahmed:2016vgl}. 
$\hat{F}^{I}_{aa}(Q^2)$ is the FF contribution, {\it i.e.}, the matrix elements of
the half-BPS or T or $\mathcal{K}$ between the on-shell state $aa$ where $a=\{\lambda,g,\phi,\chi\}$ and vacuum,
normalised by the Born contribution, which reads as 
\begin{eqnarray}
\hat F^I_{aa}(Q^2) = {\langle a(p_1),a(p_2) | \tilde {\cal O}^I | 0 \rangle \over   
                      \langle a(p_1),a(p_2) | \tilde {\cal O}^I | 0 \rangle^{(0)} }
\,,\quad \quad \quad Q^2 = (p_1+p_2)^2 \,.
\end{eqnarray}
$\tilde{\cal O}^I$ is the Fourier transform of ${\cal O}^I$ and the superscript $0$ 
indicates that it is the Born contribution. 
$\Phi^I_{aa}(z,Q^2)$ is the soft distribution function resulting from the
soft radiation and
 $\Gamma_{aa}$ are the AP kernels 
{ that can be written in terms diagonal splitting functions as given in Eq.~(\ref{diagsf})}.
\begin{strikeout}
\st{$\Gamma_{aa}$ and they contain 
only the distributions $\delta(1-z)$ and ${\cal D}_i(z)$ 
from the diagonal splitting functions $P^{(i)}_{aa}(z)$. }
\end{strikeout}
The symbol $\otimes$ denotes convolution and the ${\cal C} \exp(f(z))$ is defined by
\begin{eqnarray}
{\cal C}e^{\displaystyle f(z) }= \delta(1-z)  + {1 \over 1!} f(z)
+{1 \over 2!} f(z) \otimes f(z) + {1 \over 3!} f(z) \otimes f(z) \otimes f(z) 
+ \cdot \cdot \cdot
\end{eqnarray}
In the above, we drop all the regular terms resulting from the convolutions and keep only distributions.
In \cite{Ahmed:2016vgl},  the FFs are shown to satisfy the K+G equation 
\cite{Sudakov:1954sw,Mueller:1979ih, Collins:1980ih,Sen:1981sd}
and its solution at each order can be expressed in terms of the universal cusp ($A^I$), soft ($f^I$) and 
collinear anomalous ($B^I$) dimensions along with 
some operator dependent contributions \cite{Moch:2005tm,Ravindran:2005vv}. 
$\Delta^{I, \text{SV}}_{aa}$ is finite in the limit $\ep \rightarrow 0$, thus 
the pole structure of soft distribution function should be similar to that
of $\hat F^I_{aa}$ and $\Gamma_{aa}$. One can show that the soft distribution function
$\Phi^I_{aa}$ also satisfies a Sudakov type differential equation \cite{Ravindran:2005vv}
whose solution is straightforward to obtain:
\begin{align}
\label{phisoln}
\Phi^I_{aa} = \sum_{i=1}^{\infty}a^i\bigg(\frac{q^2(1-z)^2}{\mu_{F}^2}\bigg)^{i\epsilon/2}\bigg(\frac{1}{1-z}\bigg)\bigg[\frac{2A_{i}}{i\epsilon} -f_{i} + \overline {\cal G}^I_{ia}(\epsilon) \bigg],
\end{align}
}
where
\begin{eqnarray}
\label{softfunc}
f_1 &=& 0\,, \quad \quad \quad f_2= -28 \zeta_3
\quad \quad \quad f_3 =  {176 \over 3} \zeta_2 \zeta_3 + 192 \zeta_5.
\end{eqnarray}
We find that $\Phi^I_{aa}$ does not depend on $I$ and in addition they are identical for $a=\lambda,g,\phi$
and $\chi$.  Hence,  $\overline{\cal G}^I_{ia} = \overline {\cal G}_i$.  From the known coefficient functions, $\Delta^{I,(i),\text{SV}}$, up to two loops we can determine 
$\overline {\cal G}_i$ and they are found to be 
\begin{eqnarray}
\overline {\cal G}_1(\epsilon) &=& - 3\zeta_2\epsilon
        +{7 \over 3}\zeta_3\epsilon^2 
       - {3 \over 16}\zeta_2^2\epsilon^3
      + \left[ {31 \over 20}\zeta_5
          - {7 \over 8}\zeta_2 \zeta_3 \right]\epsilon^4\, 
\nonumber\\
&&        + \left[{49 \over 144}\zeta_3^2
          - {57 \over 640}\zeta_2^3 \right]\epsilon^5 
        +{\cal O}(\epsilon^6),
\nonumber\\
\overline {\cal G}_2(\epsilon) &=&
        4\zeta_2^2\epsilon
        +43\zeta_5\epsilon^2
        +\left[{413 \over 6}\zeta_3^2
          + {715 \over 84}\zeta_2^3 \right]\epsilon^3
\nonumber\\
&&        +\left[{9 \over 2}\zeta_7
          - {2527 \over 20}\zeta_2 \zeta_5
          + {559 \over 120}\zeta_2^2 \zeta_3\right] \epsilon^4
        +{\cal O}(\epsilon^5).
\end{eqnarray}
The above result is found to be exactly identical to $\Phi^{q}$ and $\Phi^{g}$ that appear
in the inclusive cross sections of the Drell-Yan and the Higgs productions respectively up to two loops,
after setting the Casimirs of SU(N) as $C_F=n_f=C_A$ and retaining only the LT terms.  
Our explicit computation demonstrates that the soft distribution function $\Phi$ 
contains uniform transcendental terms and in addition it obeys leading transcendentality principle.
In \cite{Ahmed:2014cla}, third order contribution to $\Phi^I$ for $I=q,g$ were obtained from \cite{Anastasiou:2014vaa}
which we use here to predict the corresponding result for $\Phi$ of ${\cal N}=4$ SYM theory after suitably adjusting the 
color factors and retaining the leading transcendental terms.  That is, we find  
\begin{eqnarray}
\label{gbar3}
f_3 &=&  {176 \over 3} \zeta_2 \zeta_3 + 192 \zeta_5.
\nonumber\\
   \overline {\cal G}_3(\epsilon) &=&
       - 4006\zeta_6
          + {536 \over 3}\zeta_3^2
          + {289192 \over 315}\zeta_2^3
        +{\cal O}(\epsilon).
\end{eqnarray}
The three-loop results for the FFs, $\hat{F}^{I}$ are already known 
\cite{Ahmed:2016vgl}, up to the same order 
the distribution parts of $\Gamma_{aa}$ (see Eq.~(\ref{diagsf}))  can be obtained by using
$A_3$ \cite{Correa:2012nk,Ahmed:2016vgl} and
$B_3$ \cite{Ahmed:2016vgl}.
Using $f_3$ 
and $\overline {\cal G}_3(\epsilon)$ from Eq.~(\ref{gbar3})
we determine $\Phi^I$ up to three loops.
Having known the form factors, soft distribution function and the AP kernels to third order,
it is now straight forward to predict the SV part cross section at third order using Eq.~(\ref{SVravi}).  
For $I={\cal K}$, we find  
\begin{eqnarray}
  \Delta_{\phi\phi}^{\mathcal{K}, (3),\text{SV}} &=&
       \left[
         - {8012 \over 3}\zeta_6
          + {13216 \over 3}\zeta_3^2
          + 480 \zeta_5
          - {992 \over 5} \zeta_2^2
          - 432 \zeta_3
          + 6512\zeta_2
          - 11552
          \right]\delta(1-z)
\nonumber\\
&&    
       + \left[
           11904\zeta_5
          - {23200 \over 3}\zeta_2 \zeta_3
          - 8736\zeta_3
          \right] {\cal D}_0
       +   \left[
       - {9856 \over 5}\zeta_2^2
+ 3712\zeta_2+9664
          \right] {\cal D}_1
      \nonumber\\ && 
       +   
           11584\zeta_3
           {\cal D}_2
+   \left[
          - 3584\zeta_2
- 3584
          \right] {\cal D}_3
       +   
           512
           {\cal D}_5.
\end{eqnarray}
and for the $I=$ half-BPS and T, we find 
\begin{eqnarray}
\label{thirdBPST}
\Delta^{I,(3),\rm{SV}}_{aa} = \Delta^{\mathcal{K},(3),\rm{SV}}_{\phi\phi} \Big |_{\rm{LT}}\,,\quad \quad \quad 
\end{eqnarray}
{where for $I=$ half-BPS, $a=\phi$ and for $I=$ T, $a=\{\lambda,g,\phi,\chi\}$}.
In addition we find that for $I=$ half-BPS, our third order prediction, Eq.~(\ref{thirdBPST}), agrees with the result 
\cite{Li:2014afw}
obtained by explicit computation. 

\section{Conclusion}
\label{conclusion}
 In this paper, we have studied the perturbative structure of ${\cal N}=4$ SYM gauge theory  
in the infrared sector and report our findings.  We achieved this by computing various 
inclusive scattering cross sections of on-shell particles belonging to this theory.  
 There are already many important perturbative results {in ${\cal N}=$ 4 SYM theory} and 
 most of them are obtained by studying on-shell scattering amplitudes.  
 These amplitudes are computed 
 in perturbation theory at leading as well as beyond the leading order in
 t'Hooft coupling, $a$. Computation of multi-loop FFs of the half-BPS operators in dimensionally
 regulated version of the theory gives perturbative coefficients such as cusp and collinear
 anomalous dimensions.  
 Unprotected operators like Konishi also demonstrate universal structure in the infrared sector
 of ${\cal N}=4$ SYM theory.  Resummed results also exist for the amplitudes and they play
 an important role in the context of AdS/CFT correspondence.                
 
 Number of computations in perturbative QCD exists, motivated to understand the 
 physics of strong interaction from the high energy colliders.  
 For example, scattering cross sections in QCD for many observables are known very precisely and 
 they are compared against the results from the experiments.
 In addition, these computations provide theoretical laboratory to 
 unravel the rich infrared structure of not only QCD but also
 a wide class of non-abelian gauge theories.  Factorisation of IR sensitive contributions
 and their universal structure in QCD amplitudes and in scattering cross sections provide
 unique opportunity to understand the infrared structure of the theory.      
 
 Motivated by these computations in QCD, we have calculated inclusive cross sections for producing
 a singlet state through the half-BPS, the energy-momentum tensor and the Konishi operators to understand 
 the soft and the collinear properties of ${\cal N}=4$ SYM theory.  By defining infrared
 safe observables in ${\cal N}=4$ SYM theory, we obtain collinear splitting functions up to second order
 in perturbation theory.  This is possible because of the factorisation of collinear
 singularities in the inclusive observables, the property that  
 infrared safe observables in QCD enjoy.  In addition, we establish the cancellation of soft divergences
 between virtual and real emission processes order by order in perturbation theory
 leaving only factorizable collinear singularities.  The former is in accordance with  
 the KLN theorem.   The systematic factorisation of collinear singularities
 and ambiguity associated with the collinear finite terms lead to RGE in the collinear sector of the theory.  The latter is governed by universal collinear
 splitting functions, analogue of AP splitting functions in perturbative QCD.  
 These splitting functions show several remarkable similarities with those of QCD.  In particular,
 only the diagonal ones contain distributions ${\cal D}_0$ and $\delta(1-z)$ with cusp and collinear anomalous
 dimensions as their coefficients, like in QCD.  In addition, several of the regular terms in $z$ are in close
 resemblance with those in QCD when the color factors of QCD are taken as $C_F=C_A=n_f=N$.       
 We find that the Mellin moments of the diagonal splitting functions in the large $j$ limit agree with
 the universal anomalous dimensions of twist-2 operators 
 when the spin $j$ becomes large. In particular, unlike~\cite{Kotikov:2003fb} we distinguish the scalar and the pseudo-scalar fields and compute the eigenvalues of the anomalous dimension matrix. We find that two of the eigenvalues coincide with the universal eigenvalues obtained in~\cite{Kotikov:2003fb} after finite shifts and the remaining two coincide with the universal ones only in the large $j$ limit. Here we wish to point out few checks on the validity of our calculation of splitting functions:
 \begin{itemize}
  \item Many of the splitting functions are calculated by considering completely different set of processes and they are found to be identical. 
 \item Our  splitting function results satisfy the identity in Eq.~(\ref{check2}).
 \item The LT terms of SV cross sections calculated in this paper matches exactly with the SM counterparts which provides third but not last non-trivial check on our computation. We elaborate further on this point below.
 \end{itemize}
 We have investigated the structure of infrared safe cross sections
 resulting after collinear factorisation.  We find that the LT terms of SV
 part of the cross sections agree with that of Drell-Yan or Higgs production cross sections in QCD when we 
 set $C_A=C_F=n_f=N$ in the latter.  This corresponds to leading transcendentality principle advocated in \cite{Kotikov:2000pm,Kotikov:2002ab,Kotikov:2004er,Kotikov:2003fb,Kotikov:2006ts}
 between the anomalous dimensions of twist-2 Wilson operators in ${\cal N}=4$ SYM theory and those of splitting functions 
 in QCD.  In addition, we find that the soft parts of the cross sections for the half-BPS, T and Konishi are all 
 identical and are independent of incoming states.   We extract the  
 soft distribution functions from inclusive cross sections and found that they are process independent,  
 namely they do not depend on the incoming states and also on the nature of singlet final state.          
 This distribution up to second order in $a$ coincides with that of Drell-Yan or Higgs production 
 when $C_A=C_F=n_f=N$ in QCD.  This is again an example for the leading transcendentality principle in the context of
 soft distribution functions in inclusive scattering cross sections.  Extending this principle to third order
 in $a$ and using the three loop FFs of the half-BPS,T and Konishi and the third order
 soft distribution function obtained from Drell-Yan or Higgs production cross sections, we have predicted 
 third order inclusive cross section $\Delta^{I,(3), \rm SV}$ for $I=$ half-BPS,T and Konishi.  Our prediction for
 the half-BPS agrees with the result obtained by explicit computation in \cite{Li:2014afw}.  $\Delta^{\text{T},(3),\rm SV}$ coincides
 identically with the half-BPS because because their three loop FFs are also identical to each other.  
 For the Konishi, the SV part of the cross section contains sub-leading
 transcendental terms unlike the case of the half-BPS or T but the leading ones coincide with those of the half-BPS and T. 
 In summary, collinear finite inclusive cross sections in ${\cal N}=4$ SYM theory provide several valuable informations
 on the perturbative IR structure of the theory.

\section*{Acknowledgement}
We thank Ajjath A.H. and Pooja Mukherjee for useful discussions. 
We acknowledge the help from our computer staff G. Srinivasan.
AC would like to thanks L. Magnea for fruitful discussions.
AC also likes to thank his parents and friends for their wonderful 
love and continuous support.

\appendix
\section{The Mellin $j$-space results for two-loop splitting functions}
\label{appA}
In the following, we list the results of two-loop splitting functions after
transforming them into Mellin $j$-space. Using Eq.~(\ref{spano}) order by order in perturbation theory and splitting function results in Eq.~(\ref{eq:expresf2l}), we obtain
\begin{eqnarray}
\label{MellinJ}
\gamma_{\phi \phi,j}^{(1)} &=& \frac{24}{j-1} + \frac{24}{j^{2}} - \frac{112}{3j} - \frac{24}{(j+1)^{2}} + \frac{40}{j+1} - \frac{16}{(j+2)^{2}}-\frac{24}{(j+2)} 
\nn  &&
+ 2\hat{Q}(j)  + \frac{8}{3}S_{1}(j-1)\, , 
\nn
\gamma_{gg,j}^{(1)}&= &-\frac{1072}{9(j-1)} - \frac{32}{j^{3}} + \frac{144}{j^{2}}+ \frac{248}{3j} + \frac{112}{(j+1)^{2}}- \frac{208}{j+1} - \frac{32}{(j+2)^{3}} + \frac{352}{3(j+2)^{2}}
\nn&&
 + \frac{2224}{9(j+2)}- 32K(j-1) + 32K(j) - 32K(j+1) + 32K(j+2) + 2\hat{Q}(j) 
 \nn&&
   + \frac{8}{3}S_{1}(j-1) \, ,
\nn
\gamma_{\lambda\lambda,j}^{(1)}&=& \frac{640}{9(j-1)} + \frac{64}{j^{3}} - \frac{192}{j^2} + \frac{8}{3j} -\frac{64}{(j+1)^2}+ \frac{128}{j+1}- \frac{256}{3(j+2)^2} - \frac{1792}{9(j+2)}  
\nn && 
- 64K(j)+ 64K(j+1)  + 2\hat{Q}(j)  + \frac{8}{3}S_{1}(j-1)\, ,
\nn
\gamma_{\lambda g,j}^{(1)}& = &\frac{640}{9(j-1)} + \frac{64}{j^{3}} - \frac{192}{j^2} - \frac{320}{(j+1)^{2}} + \frac{896}{(j+1)} + \frac{128}{(j+2)^{3}} - \frac{1792}{3(j+2)^2} 
\nn &&
 -\frac{8704}{9(j+2)}- 64K(j) + 128K(j+1) - 128K(j+2)\, ,
\nn
\gamma_{\phi g,j}^{(1)}&=& \frac{24}{j-1} + \frac{24}{j^2} - \frac{40}{j} + \frac{104}{(j+1)^{2}} - \frac{344}{(j+1)} - \frac{48}{(j+2)^{3}} + \frac{240}{(j+2)^2} + \frac{360}{(j+2)} 
\nn && 
- 48K(j+1) + 48K(j+2)\, ,
 \nn
\gamma_{g \lambda,j}^{(1)}&= &-\frac{1072}{9(j-1)} - \frac{32}{j^{3}} + \frac{144}{j^2} + \frac{80}{j}  + \frac{64}{3(j+2)^2} + \frac{352}{9(j+2)}+ \frac{32}{(j+1)^{2}} 
\nn&& 
 - 32K(j-1)+32K(j) -16K(j+1)\, ,
 \nn
\gamma_{\phi \lambda,j}^{(1)}&= &\frac{24}{j-1} + \frac{24}{j^2} - \frac{40}{j} + \frac{16}{(j+1)^{2}}- \frac{64}{(j+1)} +  \frac{32}{(j+2)^2} + \frac{80}{(j+2)} - 24K(j+1)  \, ,
\nn
\gamma_{g \phi,j}^{(1)}&= &-\frac{1072}{9(j-1)} - \frac{32}{j^{3}} + \frac{144}{j^2} + \frac{240}{3j} - \frac{16}{(j+1)^{2}} + \frac{48}{(j+1)} - \frac{32}{3(j+2)^2} - \frac{80}{9(j+2)}
 \nn&&
  -32K(j-1) + 32K(j)\, ,
  \nn
  \gamma_{\lambda \phi,j}^{(1)}&= &\frac{640}{9(j-1)} + \frac{64}{j^{3}} - \frac{192}{j^2} +\frac{64}{(j+1)^{2}} - \frac{128}{(j+1)} +  \frac{128}{3(j+2)^2} + \frac{512}{9(j+2)} 
  \nn
  &&- 64K(j)\,,
  \nn
\gamma_{\chi \phi,j}^{(1)}&= &\frac{24}{j-1} - \frac{24}{(j+1)^{2}}  + \frac{40}{(j+1)} - \frac{16}{(j+2)^2}  -\frac{24}{(j+2)} + \frac{24}{j^2} -\frac{40}{j},
  \end{eqnarray}
where
\begin{eqnarray}
\label{eqn:A2}
K(j) &= &\frac{S_{1}(j)}{j^2} + \frac{S_{2}(j)}{j} + \frac{\hat{S}_{2}(j)}{j} \, ,
\nn 
\hat{Q}(j)&=&-\frac{4}{3}S_{1}(j)+16S_{1}(j)S_{2}(j)+8S_{3}(j)-88 \hat{S}_{3}(j)+16 \hat{S}_{1,2}(j)\, ,
\nn 
S_{k}(j) & =&  \sum_{i=1}^{j} \frac{1}{i^k}\,, \quad
\hat{S}_{k}(j) = \sum _{i=1}^{j} \frac{(-1)^i}{i^k}\,, \quad
\hat{S}_{k,l}(j)= \sum_{i=1}^{j} \frac{\hat{S}_{l}(i)}{i^{k}}\, .
 \end{eqnarray}
\section{Eigenvalues of the anomalous dimension matrix}
\label{appB}
In the following, we list the expressions for the eigenvalues of the anomalous dimension matrix in Mellin $j$-space and they are found to be
\begin{align}
\lambda_1 &=\frac{8}{3j}+\frac{32}{j^3}-32K(j)-32K(j-1)+2\hat{Q}(j)+\frac{8}{3}S_1(j-1),
\nonumber\\
\lambda_2 &= \frac{8}{3j}+2\hat{Q}(j)+\frac{8}{3}S_1(j-1),
\nonumber\\
\lambda_3 &= \mathcal{E}_1+16\sqrt{\mathcal{E}_2},
\nonumber\\
\lambda_4 &= \mathcal{E}_1-16\sqrt{\mathcal{E}_2},
\end{align}
with
\begin{align}
\mathcal{E}_1 =& \frac{8}{3j}-\frac{16}{(j+2)^3}+16K(j+1)+16K(j+2) + 2\hat{Q}(j)+\frac{8}{3}S_1(j-1),
\nonumber\\
\mathcal{E}_2 =& \frac{64}{(j+1)^2}+\frac{8}{(j+1)^3}-\frac{8}{(j+1)^4} + \frac{64}{(j+2)^2}+\frac{8}{(j+2)^3}-\frac{24}{(j+2)^4}+\frac{1}{(j+2)^6}
\nonumber\\&
-\frac{128}{(j+1)(j+2)}-\frac{8}{(j+1)^2(j+2)}-\frac{8}{(j+1)(j+2)^2}-\frac{40}{(j+1)^2(j+2)^2}
\nonumber\\&
+\frac{8}{(j+1)(j+2)^3}+\frac{8}{(j+1)^2(j+2)^3} + K(j+1)\bigg[\frac{8}{j+1}-\frac{8}{j+2}+\frac{8}{(j+2)^2}
\nonumber\\&
-\frac{2}{(j+2)^3} \bigg]-K(j+2)\left[ \frac{8}{j+1}+\frac{8}{(j+1)^2}-\frac{8}{j+2}+\frac{2}{(j+2)^3}\right]
\nonumber\\&
+\Big(K(j+1)+K(j+2)\Big)^2,
\end{align}
where ${K}(j), \hat{Q}(j)$ and $S_1(j)$ can be found in Eq.~(\ref{eqn:A2}).

\bibliography{N4SF}\bibliographystyle{utphysM}

\end{document}